\pdfoutput=1
\documentclass[11pt,twoside,a4paper,cmspaper,final,collab]{cms-tdr}
\def\svnVersion{f019cae}\def\svnDate{2023/08/17}\def\cmsCernNoTag{CERN-EP-2022-172}\def\cmsCernDate{\today}\def\cmsMessage{Published in Physical Review D as \href{http://dx.doi.org/10.1103/PhysRevD.108.032008}{\doi{10.1103/PhysRevD.108.032008}.}}
\begin{document}\cmsNoteHeader{TOP-21-003}

\newcommand{\SMEFTatNLO}{\textsc{smeft@nlo}\xspace}

\newcommand{\ttH}{\ensuremath{\ttbar\PH}\xspace}
\newcommand{\ttW}{\ensuremath{\ttbar\PW}\xspace}
\newcommand{\ttZ}{\ensuremath{\ttbar\PZ}\xspace}
\newcommand{\ttgamma}{\ensuremath{\ttbar\gamma}\xspace}
\newcommand{\ttjets}{\ensuremath{\ttbar+\text{jets}}\xspace}
\newcommand{\ttbb}{\ensuremath{\ttbar+\PQb\PAQb}\xspace}
\newcommand{\ttcc}{\ensuremath{\ttbar+\PQc\PAQc}\xspace}
\newcommand{\ttLF}{\ensuremath{\ttbar+\text{LF}}\xspace}
\newcommand{\tttwob}{\ensuremath{\ttbar+2\PQb}\xspace}
\newcommand{\wjets}{\ensuremath{\PW+\text{jets}}\xspace}

\newcommand{\mtop}{\ensuremath{m_\PQt}\xspace}
\newcommand{\msd}{\ensuremath{m_\text{SD}}\xspace}
\newcommand{\hdamp}{\ensuremath{h_\text{damp}}\xspace}

\newcommand{\muR}{\ensuremath{\mu_\mathrm{R}}\xspace}
\newcommand{\muF}{\ensuremath{\mu_\mathrm{F}}\xspace}

\newcommand{\ctp}  {\ensuremath{c_{\PQt\varphi}}\xspace}
\newcommand{\ctpI} {\ensuremath{c^\mathrm{I}_{\PQt\varphi}}\xspace}
\newcommand{\cpQM} {\ensuremath{c^{-}_{\varphi\PQQ}}\xspace}
\newcommand{\cpQa} {\ensuremath{c^{3}_{\varphi\PQQ}}\xspace}
\newcommand{\cpt}  {\ensuremath{c_{\varphi\PQt}}\xspace}
\newcommand{\cptb} {\ensuremath{c_{\varphi\PQt\PQb}}\xspace}
\newcommand{\cptbI}{\ensuremath{c^\mathrm{I}_{\varphi\PQt\PQb}}\xspace}
\newcommand{\ctW}  {\ensuremath{c_{\PQt\PW}}\xspace}
\newcommand{\ctWI} {\ensuremath{c^\mathrm{I}_{\PQt\PW}}\xspace}
\newcommand{\ctZ}  {\ensuremath{c_{\PQt\PZ}}\xspace}
\newcommand{\ctZI} {\ensuremath{c^\mathrm{I}_{\PQt\PZ}}\xspace}
\newcommand{\cbW}  {\ensuremath{c_{\PQb\PW}}\xspace}
\newcommand{\cbWI} {\ensuremath{c^\mathrm{I}_{\PQb\PW}}\xspace}

\newcommand{\SW}{\ensuremath{\mathcal{S}_{\mathrm{W}}}\xspace}
\newcommand{\CW}{\ensuremath{\mathcal{C}_{\mathrm{W}}}\xspace}

\newcommand{\hc}[1]{\ensuremath{{}^\ddagger #1}}
\newcommand{\Oub}{\ensuremath{O_{\PQu\PB}^\mathrm{(ij)}}\xspace}
\newcommand{\Opud}{\ensuremath{O_{\varphi\PQu\PQd}^\mathrm{(ij)}}\xspace}
\newcommand{\Oup}{\ensuremath{O_{\PQu\varphi}^\mathrm{(ij)}}\xspace}
\newcommand{\Opq}{\ensuremath{O_{\varphi\PQq}^\mathrm{(ij)}}\xspace}
\newcommand{\Opqa}{\ensuremath{O_{\varphi\PQq}^{3\mathrm{(ij)}}}\xspace}
\newcommand{\Opu}{\ensuremath{O_{\varphi\PQu}^\mathrm{(ij)}}\xspace}
\newcommand{\OuW}{\ensuremath{O_{\PQu\PW}^\mathrm{(ij)}}\xspace}
\newcommand{\OdW}{\ensuremath{O_{\PQd\PW}^\mathrm{(ij)}}\xspace}
\newcommand{\FDF}{\ensuremath{(\varphi^\dagger i\!\overleftrightarrow{D}_{\!\!\mu}\varphi)}}
\newcommand{\FDFI}{\ensuremath{(\varphi^\dagger i\!\overleftrightarrow{D}^\mathrm{I}_{\!\!\mu}\varphi)}}

\ifthenelse{\boolean{cms@external}}{\providecommand{\cmsLeft}{upper\xspace}}{\providecommand{\cmsLeft}{left\xspace}}
\ifthenelse{\boolean{cms@external}}{\providecommand{\cmsRight}{lower\xspace}}{\providecommand{\cmsRight}{right\xspace}}

\ifthenelse{\boolean{cms@external}}{\providecommand{\cmsUpperLeft}{upper\xspace}}{\providecommand{\cmsUpperLeft}{upper left\xspace}}
\ifthenelse{\boolean{cms@external}}{\providecommand{\cmsUpperRight}{middle\xspace}}{\providecommand{\cmsUpperRight}{upper right\xspace}}

\providecommand{\cmsTable}[1]{\resizebox{\textwidth}{!}{#1}}
\newlength\cmsTabSkip\setlength{\cmsTabSkip}{1ex}

\newlength{\apptabwidth}
\ifthenelse{\boolean{cms@external}}{\setlength{\apptabwidth}{38em}}{\setlength{\apptabwidth}{33.5em}}

\newlength{\onecolwidth}
\ifthenelse{\boolean{cms@external}}{\setlength{\onecolwidth}{0.48\textwidth}}{\setlength{\onecolwidth}{0.75\textwidth}}

\newlength{\diagramwidth}
\ifthenelse{\boolean{cms@external}}{\setlength{\diagramwidth}{0.23\textwidth}}{\setlength{\diagramwidth}{0.32\textwidth}}

\cmsNoteHeader{TOP-21-003}
\title{Search for new physics using effective field theory in \texorpdfstring{13\TeV $\Pp\Pp$}{13 TeV pp} collision events that contain a top quark pair and a boosted \texorpdfstring{\PZ}{Z} or Higgs boson}

\author*[BAY]{B.~Caraway}

\date{\today}

\abstract{
A data sample containing top quark pairs (\ttbar) produced in association with a Lorentz-boosted \PZ or Higgs boson is used to search for signs of new physics using effective field theory.  The data correspond to an integrated luminosity of 138\fbinv of proton-proton collisions produced at a center-of-mass energy of 13\TeV at the LHC and collected by the CMS experiment.  Selected events contain a single lepton and hadronic jets, including two identified with the decay of bottom quarks, plus an additional large-radius jet with high transverse momentum identified as a \PZ or Higgs boson decaying to a bottom quark pair.  Machine learning techniques are employed to discriminate between \ttZ or \ttH events and events from background processes, which are dominated by \ttjets production.  No indications of new physics are observed.  The signal strengths of boosted \ttZ and \ttH production are measured, and upper limits are placed on the \ttZ and \ttH differential cross sections as functions of the \PZ or Higgs boson transverse momentum.  The effects of new physics are probed using a framework in which the standard model is considered to be the low-energy effective field theory of a higher energy scale theory.  Eight possible dimension-six operators are added to the standard model Lagrangian and their corresponding coefficients are constrained via fits to the data.
}
\hypersetup{
pdfauthor={CMS Collaboration},
pdftitle={Search for new physics using effective field theory in 13 TeV pp collision events that contain a top quark pair and a boosted Z or Higgs boson},
pdfsubject={CMS},
pdfkeywords={CMS, top, Z boson, Higgs, EFT, boosted}}

\maketitle

\section{Introduction}
\label{sec:intro}
The standard model (SM) of particle physics successfully describes a vast range of subatomic phenomena with outstanding precision.
However, it cannot explain the existence of dark matter~\cite{Rubin:1970zza,Aghanim:2018eyx,Clowe:2003tk} and faces other difficulties,
such as the fine-tuning or hierarchy problem~\cite{Kaul:1981hi,Barbieri:1987fn}.
This indicates that the SM is incomplete and may be only a low-energy approximation, which is valid at the energies currently accessible to experiments, to a more fundamental theory.
Null results from direct searches for new particles predicted by various models of physics beyond the SM (BSM) suggest that such new particles may be too massive to be produced at the CERN LHC.

The signatures of BSM physics arising from high-mass particles, which may not be directly accessible at the LHC, could appear indirectly in the form of deviations from the SM predictions for production of SM particles.
All possible high-energy-scale deviations from the SM can be described at low energy scales by the SM effective field theory (EFT)~\cite{Buchmuller:1985jz,Grzadkowski:2010es,Falkowski:2019tft}.
This EFT extends the SM Lagrangian $\mathcal{L}_\text{SM}$ with operators $\mathcal{O}_\mathrm{i}^{(d)}$ of dimension $d > 4$ and
effective couplings $c_\mathrm{i}^{(d)}$, known as Wilson coefficients (WCs), which are suppressed by appropriate powers of the EFT energy scale $\Lambda$, leading to an EFT Langrangian of:
\begin{equation*}
\mathcal{L}_\text{eff} = \mathcal{L}_\text{SM} + \sum_{d,\mathrm{i}} \frac{c_\mathrm{i}^{(d)}}{\Lambda^{d-4}} \mathcal{O}_\mathrm{i}^{(d)}.
\end{equation*}
Because of the $1 / \Lambda^{d-4}$ suppression of non-SM terms, the lowest-dimension operators are the most likely to have measurable effects.
Dimension-five operators produce lepton-number violation~\cite{Degrande:2012wf,Kobach:2016ami}, which is strongly constrained~\cite{PDG2020}, so this analysis only considers dimension-six operators~\cite{AguilarSaavedra:2018nen}.

The impact of these operators may be detectable in a wide variety of experimental observables.
This analysis searches for BSM effects on the production of a pair of top quarks (\ttbar) in association with a \PZ boson or Higgs boson (\PH).
These processes are particularly interesting, as they provide direct access to the couplings of the top quark to the \PZ and Higgs bosons.
Any deviations of these couplings from their SM values might imply the existence of BSM effects in the electroweak symmetry breaking mechanism,
which could be probed in an EFT framework.

\begin{figure}[h]
  \centering
	\hspace*{\fill}
    \includegraphics[width=\diagramwidth]{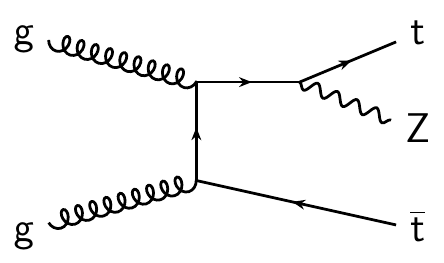}
	\hspace*{\fill}
    \includegraphics[width=\diagramwidth]{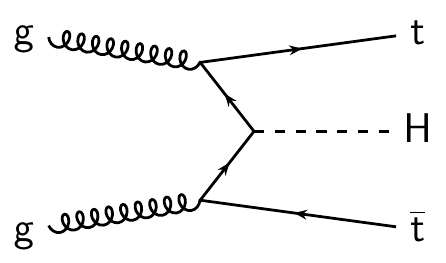}
	\hspace*{\fill}
	\caption{Examples of tree-level Feynman diagrams for the \ttZ (left) and \ttH (right) production processes.}
	\label{fig:diagrams}
\end{figure}

The \ttZ and \ttH processes (Fig.~\ref{fig:diagrams}) are challenging to measure with high precision because of their low production cross sections.
However, with the large amount of data delivered by the LHC, the ATLAS and CMS Collaborations have measured the \ttZ{}~\cite{Aaboud:2019njj,ATLAS:2021fzm,Sirunyan:2017uzs,CMS-TOP-18-009} and \ttH~\cite{Aaboud:2018urx,ATLAS:2020ior,ATLAS:2021qou,Sirunyan:2018hoz,CMS:2018sah,CMS:2018hnq,CMS:2020cga,CMS:2020mpn} cross sections with a relative uncertainty of 8 and 20\%, respectively.
Many of these measurements, especially of \ttZ, have been performed in final states containing multiple charged leptons.
Additionally, an extensive exploration of EFT effects in collisions producing \ttbar in association with a \PW, \PZ, or Higgs boson has been performed by the CMS Collaboration, also in final states containing multiple charged leptons~\cite{CMS-TOP-19-001}.
Other recent analyses have constrained the WCs corresponding to specific EFT operators using differential and inclusive \ttZ and \ttgamma measurements~\cite{Aaboud:2019njj,CMS-TOP-18-009,CMS-TOP-18-010,CMS-TOP-21-001,CMS-TOP-21-004}.

\begin{figure}
  \centering
    \includegraphics[width=0.48\textwidth]{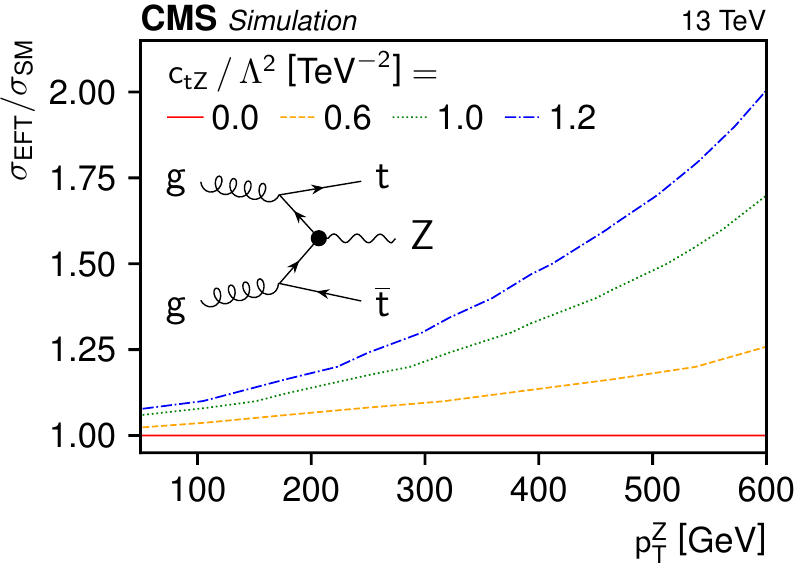}
    \includegraphics[width=0.48\textwidth]{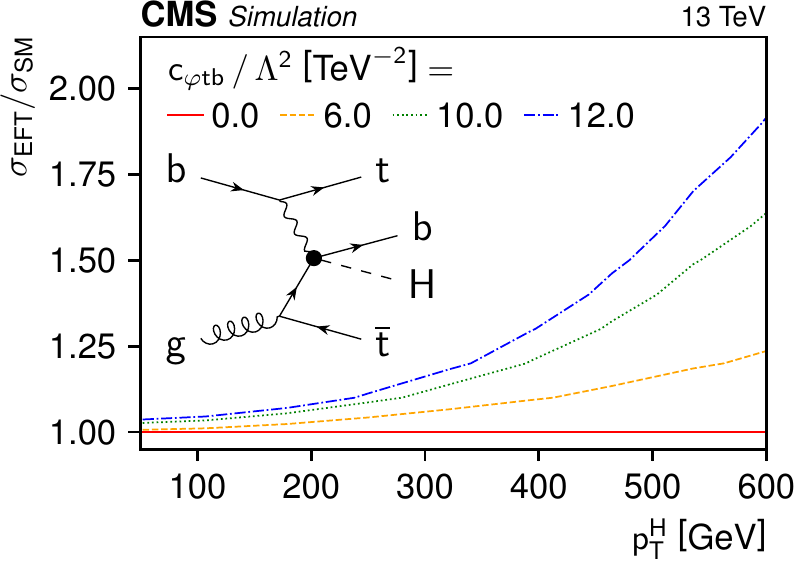}
	\caption{The \ttZ (\cmsLeft) and \ttH (\cmsRight) cross sections in the SM EFT, as ratios to the corresponding SM cross sections, as functions of $\ctZ/{\Lambda}^{2}$ and the \PZ boson \pt (\cmsLeft), and $\cptb/{\Lambda}^{2}$ and the Higgs boson \pt (\cmsRight), where \ctZ and \cptb are the WCs for the EFT operators \Oub and \Opud, respectively~\cite{Grzadkowski:2010es}.
    Example Feynman diagrams, in which the vertices affected by \ctZ and \cptb are marked by a filled dot, are also displayed.}
\label{fig:analysis:eftimpact}
\end{figure}

The fraction of \ttZ and \ttH events whose decay products include a single lepton is much larger than for multiple leptons, but the large background from \ttbar produced in association with multiple jets from quantum chromodynamic (QCD) processes (\ttjets) makes measurements in single-lepton final states extremely challenging.
However, at high \PZ or Higgs boson transverse momentum (\pt), where EFT effects are generally more pronounced, as shown in Fig.~\ref{fig:analysis:eftimpact}, we can utilize jet substructure techniques to identify $\PZ/\PH\to\bbbar$ decays that are reconstructed as a single merged jet, and suppress the \ttjets background.
In this analysis, we probe EFT effects in \ttZ and \ttH events with a single lepton from the decay of one of the top quarks and a high-\pt \PZ or Higgs boson decaying to $\bbbar$.

We identify high-\pt \PZ or Higgs bosons decaying to $\bbbar$ using a large-radius jet, described below, which
is identified using a machine learning technique as being consistent with the decay of a heavy boson to $\bbbar$.
The top quark pair is identified using the combination of a single electron or muon,
moderate missing transverse momentum (\ptmiss, defined below),
at least two small-radius jets identified as originating from bottom quarks,
which must be well separated from the large-radius \PZ or Higgs boson jet,
and additional small-radius jets to account for the hadronically decaying \PW boson.
A neural network is used to separate signal-like events from the dominant \ttjets background,
and the reconstructed mass of the large-radius jet serves to distinguish \ttZ and \ttH events from each other, as well as from the background.
A global fit to the data measures the signal strengths of the \ttZ and \ttH processes.
We also place upper limits on the \ttZ and \ttH differential cross sections as functions of the \pt of the \PZ boson ($\pt^{\PZ}$) and of the Higgs boson ($\pt^{\PH}$),
and constrain the WCs corresponding to EFT operators that most strongly affect boosted \ttZ and \ttH topologies.

In this paper, we briefly describe the CMS detector in Section~\ref{sec:detector} and the Monte Carlo simulation of the signals and backgrounds in Section~\ref{sec:samples}.
Sections~\ref{sec:reconstruction} and \ref{sec:selection} discuss the reconstruction of particles and selection of events, respectively, while Section~\ref{sec:discrimination} and Appendix~\ref{app:mva_vars} describe the neural network used to separate signal-like events from the backgrounds.
In Sections~\ref{sec:systematics} and \ref{sec:results}, we detail the sources of systematic uncertainty in our measurements and present the results of those measurements.
Lastly, a summary is provided in Section~\ref{sec:summary}.
Tabulated results are provided in the HEPData record for this analysis~\cite{hepdata}.

\section{The CMS detector}
\label{sec:detector}
The CMS detector is a general-purpose particle detector at the LHC.
It is centered on a luminous region where the LHC beams interact and is roughly symmetric under rotations around the beam direction.
The inner detector is contained within a 3.8\unit{T} magnetic field produced by a 6\unit{m} internal diameter solenoid,
and is composed of silicon pixel and silicon strip tracking detectors as well as a lead tungstate crystal electromagnetic calorimeter (ECAL) and a brass and scintillator hadron calorimeter.
Each component of the inner detector is in turn divided into a cylindrical barrel section and two endcap sections.
The magnetic flux from the solenoid is guided through a steel return yoke.
Gas-ionization detectors embedded in the flux-return yoke are used to detect muons.

Events are collected using a two-tier trigger system: a hardware-based level-1 trigger~\cite{CMS:2020cmk} and a software-based high-level trigger~\cite{CMS_trigger}.
The CMS detector is described in more detail along with the coordinate system and basic kinematic variables in Ref.~\cite{CMS}.

\section{Data and simulation samples}
\label{sec:samples}
This analysis is performed using proton-proton ($\Pp\Pp$) collision data recorded at $\sqrt{s}=13\TeV$ at the LHC corresponding to an integrated luminosity of 138\fbinv, of which 36.3, 41.5, and 59.7\fbinv were recorded in 2016, 2017, and 2018, respectively~\cite{CMS-LUM-17-003,CMS-PAS-LUM-17-004,CMS-PAS-LUM-18-002}.
The data and simulation samples from these three years are treated separately to account for variations in the detector and LHC conditions.
The data were recorded using combinations of triggers that require a single electron or muon candidate~\cite{CMS_trigger}.
The minimum trigger \pt thresholds range from 27 to 32\GeV for electrons and from 24 to 27\GeV for muons, depending on the data-taking period.

Monte Carlo (MC) simulation is used to model signal and background events.
For the \ttZ and \ttH signals, and for the subset of \ttbar events containing at least one \PQb hadron that does not originate from a top quark decay (\ttbb),
two sets of MC samples are used: one set is generated at next-to-leading order (NLO) in perturbative QCD,
and the other set is generated at leading order (LO) in QCD with weights corresponding to a range of WC values, representing various BSM scenarios in the SM EFT.
The NLO samples (referred to as ``SM samples'') are used to model the expected \ttZ, \ttH, and \ttbb contributions, while the LO samples with EFT weights (``EFT samples'') are used to evaluate EFT effects relative to the SM predictions.
The {\MGvATNLO}\,v2.6.0 program~\cite{Alwall:2014hca} is employed to generate the \ttZ SM signal samples, while the \ttH SM signal samples are generated with the {\POWHEG}\,v2.0~\cite{Nason:2004rx,Frixione:2007vw,Alioli:2010xd,Hartanto:2015uka} program.
The SM \ttZ signal sample is normalized to a production cross section of 0.86\unit{pb}, which is computed at NLO with next-to-next-to-leading logarithmic (NNLL) resummation in QCD~\cite{Kulesza:2018tqz}.
Similarly, the SM \ttH signal sample is normalized to a production cross section of 0.507\unit{pb}, which is computed at NLO in QCD and electroweak corrections~\cite{deFlorian:2016spz} when its SM predictions are presented.

The dominant background arises from \ttbar production.
Simulated \ttbar events are separated into three flavor categories.
These categories are based on the flavor of additional jets at the particle level that do not originate from the top quark decays and
that have $\pt>20\GeV$ and pseudorapidity $\abs{\eta}<2.4$:
\ttbb consists of events with at least one \PQb hadron that does not originate from a top quark decay;
\ttcc consists of events with at least one additional jet that originates from one or more \PQc hadrons and no additional jets from \PQb hadrons;
and \ttLF (light flavor) includes all remaining events.
Within the \ttbb category, the subset of events with two additional \PQb hadrons that are sufficiently collinear to produce a single jet is denoted by \tttwob~\cite{CMS:2018sah,CMS:2018hnq},
and is subject to additional theoretical uncertainties~\cite{ttbb:2020_frag_hf}.
The \ttbb, \ttcc, and \ttLF events are collectively referred to as \ttjets events.

The \ttbb events, which are the most critical background, are simulated at NLO in the four-flavor scheme (4FS), following the studies in Refs.~\cite{Jezo:2018yaf,Buccioni:2019plc}.
This is done using the dedicated \POWHEG settings presented in Ref.~\cite{Jezo:2018yaf}, except that the values of the factorization and renormalization scales have been reduced by one half to approximate the effects of higher-order corrections following Ref.~\cite{Buccioni:2019plc}.
The \ttcc and \ttLF contributions are simulated using \ttbar events generated with \POWHEG at NLO in the five-flavor scheme (5FS)~\cite{Nason:2004rx,Frixione:2007vw,Alioli:2010xd,Frixione:2007nw}.
The top quark \pt spectrum in simulation is reweighted to match the prediction at next-to-next-to-leading order (NNLO) in QCD and NLO in electroweak corrections~\cite{Czakon:2017wor}.
The total yield of the \ttbar events is scaled to the inclusive \ttbar cross section of 832\unit{pb}, which corresponds to the NNLO+NNLL prediction~\cite{Beneke:2011mq, Cacciari:2011hy, Baernreuther:2012ws,Czakon:2012zr,Czakon:2012pz,Czakon:2013goa,Czakon:2011xx}.
The fractional contributions of the \ttbb, \ttcc, and \ttLF processes are set according to the 5FS \ttbar simulation, and the final \ttbb yield is a free parameter in the fit to the data.

The {\MGvATNLO} program~\cite{Alwall:2014hca} (v2.2.2 for 2016 and v2.4.2 for 2017--2018)
is used to generate the \ttW, \ttgamma, \ttbar{}\ttbar{}, $s$-channel single top quark, and \PQt{}\PZ{}\PQq background samples at NLO in QCD.
The same program is also used to generate the \wjets, Drell--Yan (DY)${}+\text{jets}$, $\PQt\PH\PQq$, and $\PQt\PH\PW$ background samples at LO in QCD.
In addition to the \ttbar samples described above, the $t$-channel single top quark and $\PQt\PW$ background samples are generated with the {\POWHEG}\,v2.0~\cite{Nason:2004rx,Frixione:2007vw,Alioli:2010xd,Frixione:2007nw,Alioli:2009je,Re:2010bp} program at NLO in QCD.
The Higgs boson and top quark masses are taken to be 125 and 172.5\GeV, respectively.

The simulated backgrounds other than \ttbar are also normalized to their predicted cross sections, which are
taken from theoretical calculations at NLO or NNLO in QCD~\cite{Kulesza:2018tqz,Gavin:2012sy,Li:2012wna,Kidonakis:2010ux,Aliev:2010zk,Kant:2014oha}.

{\tolerance=520
All of the simulated samples use {\PYTHIA}\,8.226~(8.230) to describe parton showering and hadronization for 2016 (2017 and 2018).
The samples that are simulated at NLO with {\MGvATNLO} use the FxFx~\cite{Frederix:2012ps} scheme for matching partons from the matrix element calculation to those from parton showers.
The samples simulated at LO use the MLM~\cite{Mangano:2006rw} scheme for the same purpose.
The parameters for the underlying-event description correspond to the CP5 tune~\cite{Sirunyan:2019dfx} except for the background samples generated with {\MGvATNLO} for 2016,
for which the CUETP8M1~\cite{Khachatryan:2015pea} tune is used.
The samples using the CP5 tune employ the NNPDF3.1 NNLO~\cite{Ball:2017nwa} parton distribution functions (PDFs),
whereas samples using the CUETP8M1 tune employ the NNPDF3.0 LO~\cite{Ball:2014uwa} PDFs.
\par}

The simulated events are processed through a \GEANTfour-based~\cite{Agostinelli:2002hh} simulation of the CMS detector,
including the effects of additional $\Pp\Pp$ interactions per bunch crossing, referred to as pileup.
The events are reweighted to match the estimated pileup distribution in data, based on the measured instantaneous luminosity.

\subsection{Simulation of EFT samples}
\label{sec:eft_samples}

This analysis considers eight dimension-six EFT operators in the Warsaw basis~\cite{Grzadkowski:2010es} that have a large impact on the \ttZ and \ttH signal processes and minimal impacts on the background predictions.
Table~\ref{tab:eftOperators} lists the eight operators and corresponding WCs.
All couplings are restricted to involve only third-generation quarks because first- and second-generation
couplings have been strongly constrained by other analyses and because the \ttZ and \ttH processes are
uniquely sensitive to third-generation couplings.
The operators that also require their Hermitian conjugate term in the Lagrangian (marked with $\ddagger$ in Table~\ref{tab:eftOperators}) can have complex WCs;
however, imaginary
coefficients lead to charge-conjugation and parity violation, and are generally already constrained~\cite{AguilarSaavedra:2018nen}.
Therefore, following a similar strategy to that used in an earlier CMS analysis~\cite{CMS-TOP-19-001},
only the eight real-component WCs in Table~\ref{tab:eftOperators} are considered in this analysis.

\begin{table}
  \topcaption{The set of EFT operators considered in this analysis that affect the \ttZ and \ttH processes at order $1/\Lambda^2$.
    The couplings are restricted to involve only third-generation quarks.
	The symbol $\gamma^\mu$ denotes the Dirac matrices and $\sigma^{\mu\nu} \equiv i \left(\gamma^\mu \gamma^\nu - g^{\mu \nu}\right)$, where $g^{\mu \nu}$ is the metric tensor and $\tau^\mathrm{I}$ are the Pauli matrices.
    The field $\varphi$ is the Higgs boson doublet and $\tilde{\varphi}^\mathrm{j} \equiv \varepsilon_\mathrm{jk} \left(\varphi^\mathrm{k}\right)^*$, where $\varepsilon_\mathrm{jk}$ is the Levi--Civita symbol and $\varepsilon_{12} = +1$.
    The quark doublet and the right-handed quark singlets are represented by \PQq, \PQu, and \PQd, respectively.
    The quantities
    $\protect\FDF \equiv \varphi^\dagger(iD_\mu\varphi)-(iD_\mu\varphi^\dagger)\varphi$ and
	$\protect\FDFI \equiv \varphi^{\dagger}\tau^\mathrm{I}(iD_\mu\varphi)-(iD_\mu\varphi^\dagger)\tau^\mathrm{I}\varphi$,
    where $D_{\mu}$ is the covariant derivative.
	The symbols $\PW_{\mu\nu}^\mathrm{I}$ and $\PB_{\mu\nu}$ are the field strength tensors for the weak isospin and weak hypercharge gauge fields.
    The abbreviations \SW and \CW denote the sine and cosine of the weak mixing angle in the unitary gauge, respectively.
    The operators marked with the $\ddagger$ symbol also require their Hermitian conjugate in the Lagrangian.
  }
\centering
\begin{scotch}{lll}
Operator                        & Definition                                                                                                & WC                                     \\
\hline
	\hc{\Oup} \rule{0pt}{2.6ex} & $\PAQq_\mathrm{i} \PQu_\mathrm{j}\tilde\varphi\: (\varphi^{\dagger}\varphi)$                              & $\ctp + i \ctpI$                       \\
	\Opq                        & $\FDF (\PAQq_\mathrm{i}\gamma^\mu \PQq_\mathrm{j})$                                                       & $\cpQM + \cpQa$                        \\
	\Opqa                       & $\FDFI (\PAQq_\mathrm{i}\gamma^\mu\tau^\mathrm{I} \PQq_\mathrm{j})$                                       & $\cpQa$                                \\
	\Opu                        & $\FDF (\PAQu_\mathrm{i}\gamma^\mu \PQu_\mathrm{j})$                                                       & $\cpt$                                 \\
	\hc{\Opud}                  & $(\tilde\varphi^\dagger iD_\mu\varphi)(\PAQu_\mathrm{i}\gamma^\mu \PQd_\mathrm{j})$                       & $\cptb + i \cptbI$                     \\
	\hc{\OuW}                   & $(\PAQq_\mathrm{i}\sigma^{\mu\nu}\tau^\mathrm{I}\PQu_\mathrm{j})\:\tilde{\varphi}\PW_{\mu\nu}^\mathrm{I}$ & $\ctW + i \ctWI$                       \\
	\hc{\OdW}                   & $(\PAQq_\mathrm{i}\sigma^{\mu\nu}\tau^\mathrm{I}\PQd_\mathrm{j})\:{\varphi} \PW_{\mu\nu}^\mathrm{I}$      & $\cbW + i \cbWI$                       \\
\ifthenelse{\boolean{cms@external}}{
	\multirow{2}{*}{\hc{\Oub}}  & \multirow{2}{*}{$(\PAQq_\mathrm{i}\sigma^{\mu\nu}\PQu_\mathrm{j})\:\tilde{\varphi}\PB_{\mu\nu}$}          & $\frac{\CW}{\SW} (\ctW + i \ctWI)$     \\
	                            &                                                                                                           & $\,{}- \frac{1}{\SW} (\ctZ + i \ctZI)$ \\
}{
	\hc{\Oub}                   & $(\PAQq_\mathrm{i}\sigma^{\mu\nu}\PQu_\mathrm{j})\:\tilde{\varphi}\PB_{\mu\nu}$                           & $\frac{\CW}{\SW} (\ctW + i \ctWI) - \frac{1}{\SW} (\ctZ + i \ctZI)$ \\
}
\end{scotch}
\label{tab:eftOperators}
\end{table}

Samples with EFT effects are generated for the \ttZ and \ttH signal processes and the \ttbb background
using the {\MGvATNLO}\,v2.6.5 event generator at LO
and the NNPDF3.1 NNLO PDF set, following a similar approach to that outlined in Ref.~\cite{AguilarSaavedra:2018nen}.
The EFT effects on the decay of the top quark and the \PZ and Higgs bosons are expected to be negligible, and are not considered.
As is done for the SM samples,
the EFT samples are processed through a \GEANTfour-based~\cite{Agostinelli:2002hh} simulation of the CMS detector,
and event reconstruction is performed in the same manner as for the collision data.
Samples of \ttLF and \ttcc background events are also generated with EFT effects, and no observable variations are seen for the eight operators considered.
The \ttbb sample shows a nonnegligible EFT effect when varying \cbW, which is accounted for in the analysis.
The \ttZ and \ttH signal samples are generated with up to one extra parton in the final state.
This improves the precision of the simulation, bringing it closer to NLO in QCD~\cite{eftlojvsnol2021,CMS-TOP-19-001}.
The EFT effects from this approach are found to be compatible with those from the \SMEFTatNLO model~\cite{Degrande:2020evl}
for the dimension-six operators included in \SMEFTatNLO~\cite{eftlojvsnol2021}.

In the EFT samples, the effects of varying the eight WCs are computed as a weighting factor for each simulated event.
The weight $w_\text{EFT}$ for each event is parameterized as a quadratic function of the eight WCs:
\begin{equation}
	w_\text{EFT} = s_{0} + \sum_\mathrm{i} s_\mathrm{1i} \frac{c_\mathrm{i}}{\Lambda^2} + \sum_\mathrm{i, j} s_\mathrm{2ij} \frac{c_\mathrm{i}}{\Lambda^2} \frac{c_\mathrm{j}}{\Lambda^2},
\label{eq:quad_wgt_eqn}
\end{equation}
where $s_{0}$ is the weight arising only from SM processes, subscripts $\mathrm{i}$ and $\mathrm{j}$ are indices for WCs under consideration, $s_\mathrm{1i}$ is the coefficient for the term arising from the interference between one EFT operator and the SM, and $s_\mathrm{2ij}$ is the coefficient for the term arising from interference between two EFT operators (for $\mathrm{i} \neq \mathrm{j}$) or for the term arising solely from one EFT operator (for $\mathrm{i} = \mathrm{j}$).
In the SM, where all WCs are zero, $w_\text{EFT}$ is equal to $s_{0}$.

\section{Event reconstruction}
\label{sec:reconstruction}
Collisions in the CMS detector are reconstructed using the particle-flow (PF) algorithm~\cite{PF}.
This algorithm exploits information from every subsystem of the detector and produces candidate reconstructed objects (PF candidates) in several categories: charged and neutral hadrons, photons, electrons, and muons.
The PF candidates are used, in various combinations, to reconstruct higher-level objects such as interaction vertices, hadronic jets, and the \ptvecmiss.
This section describes the reconstruction of the objects that are used in this analysis.

The \ptvecmiss vector, which is used to estimate the summed \ptvec of neutrinos in the event,
is defined as the negative vector sum of the \ptvec of all PF candidates in the event~\cite{CMS_MET}.
Its magnitude is denoted by \ptmiss.

An interaction vertex is a point from which multiple tracks in one event originate.
The primary vertex (PV) is taken to be the interaction vertex corresponding to the hardest scattering in the event, evaluated using tracking information alone, as described in Section~9.4.1 of Ref.~\cite{CMS-TDR-15-02}.
All selected objects in this analysis must be associated with the PV.

Electron candidates are reconstructed and identified by combining information from the ECAL and the tracking system~\cite{CMS_electron}, and
they are required to have ${\abs{\eta} < 2.5}$ and ${\pt > 30}$ (35)\GeV for 2016 (2017 and 2018).
The variation of the electron candidate \pt requirement is related to a change in the electron trigger between 2016 and 2017.
Muon candidates are reconstructed as tracks in the silicon tracking system consistent with measurements in the muon system, and associated with calorimeter deposits compatible with the muon hypothesis~\cite{CMS_muon}.
Muon candidates are required to have ${\abs{\eta} < 2.4}$ and ${\pt > 30\GeV}$.
Electron and muon candidates are required to be isolated from surrounding hadronic activity and to be consistent with an origin at the PV.

The electron and muon isolation requirement is based on the mini-isolation variable $I_\text{mini}$, which is defined as the scalar sum of the \pt of all charged hadron, neutral hadron, and photon PF candidates within a variably sized cone around the electron or muon candidate direction~\cite{CMS:2016krz}.
The cone radius $\DR = \smash[b]{\sqrt{ \left(\Delta\eta\right)^2 + \left(\Delta\phi\right)^2}}$ is
10\GeV divided by the lepton \pt within the range $50 < \pt < 200\GeV$.
For leptons with $\pt < 50\GeV$ (${>}200\GeV$), we use $\DR = 0.2$ (0.05).
The mini-isolation variable is corrected for contributions from pileup using estimates
of the charged and neutral pileup energy inside the cone~\cite{CMS_muon,CMS_electron}.
Electron and muon candidates are required to have ${I_\text{mini} / \pt < 0.2}$ and 0.1, respectively.
The overall trigger efficiency, measured in a control region requiring one electron and one muon, is around 86\% (87\%) for events with an electron (muon) that satisfies these selection criteria.

Two types of jets, small-radius (``AK4'') and large-radius (``AK8''), are used in this analysis.
Both are reconstructed by clustering charged and neutral PF candidates using the anti-\kt jet
finding algorithm~\cite{Cacciari:2008gp,Cacciari:2011ma}, with distance parameters of 0.4 and 0.8, respectively.

The AK4 jets use charged-hadron subtraction~\cite{CMS-PU-mitigation} for pileup mitigation and are required to have ${\pt > 30\GeV}$ and ${\abs{\eta} < 2.4}$.
The AK8 jets use the pileup per particle identification algorithm~\cite{Bertolini:2014,CMS-PU-mitigation} to reduce the effects of pileup and are required to have ${\pt > 200\GeV}$, ${\abs{\eta} < 2.4}$, and an invariant mass between 50 and 200\GeV.
The invariant mass of each AK8 jet, \msd, is calculated using the soft-drop algorithm~\cite{Dasgupta:2013ihk,Larkoski:2014wba} with an angular exponent ${\beta = 0}$ and soft cutoff threshold ${z_\text{cut} = 0.1}$. This algorithm systematically removes soft and collinear radiation from the jet.  Its performance in collision data is validated in Ref.~\cite{JME-18-002}.

Jet quality criteria~\cite{CMS-PU-mitigation} are imposed to eliminate jets from spurious sources such as electronics noise.
The jet energies are corrected for residual pileup effects~\cite{Cacciari:2007fd}
and the detector response as a function of \pt and $\eta$~\cite{CMS_JES}.
All jets are required to be separated from the selected electron or muon candidate by ${\DR > 0.4}$ or 0.8 for AK4 and AK8 jets, respectively.
Selected AK4 jets may overlap spatially with selected AK8 jets, and may use the same PF candidates.

Jets originating from the hadronization of bottom quarks (\PQb jets) are identified (\PQb tagged) using a secondary-vertex algorithm based on a deep neural network~(\textsc{DeepCSV})~\cite{DeepCSV}.
The ``medium'' working point of this algorithm is used, which provides a tagging efficiency of 68\% for typical \PQb jets from top quark decays~\cite{DeepCSV}.
The corresponding misidentification probability for light-flavor jets, which contain only light-flavor hadrons and originate from gluons and up, down, and strange quarks, is 1\%, and for charm quark jets (\PQc jets) it is 12\%~\cite{DeepCSV}.

When a \PZ or Higgs boson is produced with a large Lorentz boost, its decay products will be collimated and may be reconstructed as a single AK8 jet.
We employ a multivariate discriminant to identify AK8 jets containing the decay products from a $\bbbar$ pair.
This discriminant is based on a dedicated deep neural network algorithm, the \textsc{DeepAK8} $\bbbar$ tagger~\cite{JME-18-002}, which is uncorrelated with \msd.
By combining the tagger output and \msd, we can identify jets consistent with $\PZ\to\bbbar$ and $\PH\to\bbbar$ decays and use the \msd regions away from the \PZ and Higgs boson masses as background-enriched control regions.
The $\bbbar$ tagger produces a score between 0 and 1.
The selected AK8 jet with the largest score in each event is identified
as the \PZ or Higgs boson candidate, provided that the score is greater than 0.8.
This $\bbbar$ tagger score requirement together with the \msd requirement produces a $\bbbar$ tagging
efficiency of 45--70\% for $\PZ\to\bbbar$ decays and 25--60\% for $\PH\to\bbbar$ decays, depending on the \pt of the AK8 jet.
These efficiencies in \ttZ and \ttH events are affected by contamination from top quark decay products, which is less prominent in \ttZ events than in \ttH events because the \PZ boson tends to be better separated from the top quarks.
The misidentification rate is roughly 2.5\% for jets from QCD multijet events.

\section{Event selection}
\label{sec:selection}
This analysis targets \ttZ or \ttH production, where the
\ttbar decay products include an electron or muon and a neutrino, plus two \PQb quarks and two
other quarks, and the \PZ or Higgs boson decays to a $\bbbar$ pair.
We select events containing exactly one candidate electron or muon, ${\ptmiss > 20\GeV}$,
five or more AK4 jets, and a \PZ or Higgs boson candidate AK8 jet.
The electron or muon and the \ptmiss are expected to arise from the leptonic decay of a \PW boson.
At least one (and sometimes two) of the AK4 jets spatially overlap the AK8 jet because the AK4
and AK8 jet clustering algorithms run independently with no unambiguous way to distinguish AK4
and AK8 jets that share some or all of of their constituent particles.
Three or four AK4 jets, corresponding to the hadrons from the \ttbar decay products, are expected to be well separated from the AK8 jet.
At least two AK4 jets must be \PQb tagged
and separated from the \PZ or Higgs boson candidate jet by $\DR > 0.8$.
These requirements are summarized in Table~\ref{tab:selection}.

\begin{table*}
	\topcaption{Summary of the reconstructed object and event selection requirements.}
	\label{tab:selection}
	\centering
	\cmsTable{
	\begin{scotch}{ll}
		Missing transverse momentum                      & $\ptmiss > 20\GeV$ \\ [\cmsTabSkip]

		\multirow{4}{*}{${=} 1$ electron or muon}        & $\pt(\Pe) > 30$ (35\GeV) in 2016 (2017 and 2018) \\
		                                                 & $\pt(\PGm) > 30\GeV$ \\
		                                                 & $\abs{\eta(\Pe)} < 2.5$, $\abs{\eta(\PGm)} < 2.4$ \\
														 & $I_\text{mini}(\Pe) / \pt(\Pe) < 0.2$, $I_\text{mini}(\PGm) / \pt(\PGm) < 0.1$ \\ [\cmsTabSkip]

		\multirow{2}{*}{${\geq} 1$ AK8 jet}              & $\pt > 200\GeV$, $\abs{\eta} < 2.4$ \\
		                                                 & $50 < \msd < 200\GeV$ \\ [\cmsTabSkip]

		${=} 1$ \PZ or Higgs boson candidate AK8 jet     & Highest $\bbbar$ tagger score $({>}0.8)$ \\ [\cmsTabSkip]

		${\geq} 5$ AK4 jets (may overlap AK8 jets)       & $\pt > 30\GeV$, $\abs{\eta} < 2.4$  \\ [\cmsTabSkip]

		\multirow{2}{*}{${\geq} 2$ \PQb-tagged AK4 jets} & Satisfy medium \textsc{DeepCSV} \PQb tag requirements  \\
														 & $\DR(\text{\PQb-tagged AK4 jet}, \text{\PZ or Higgs boson candidate}) > 0.8$ \\ [\cmsTabSkip]

	\end{scotch}
	}
\end{table*}

To remove events with spurious \ptmiss, we use the event filters described in Ref.~\cite{CMS_MET}.
In addition, events in which the lepton candidate and a second lepton of the same flavor but opposite charge have an
invariant mass less than 12\GeV are removed from the data set to eliminate leptons from \PJGy or \PGU meson decay.
This requirement, in combination with the \ptmiss threshold, is effective at significantly reducing the nonprompt-lepton background.
The \pt, isolation, and identification requirements are relaxed for the second lepton,
because it typically has lower momentum than the primary lepton candidate.
Lastly, we remove events that contain an electron PF candidate with $\pt > 20\GeV$, $-3 < \eta < -1.4$, and $-1.57 < \phi < -0.87$ during the latter part of 2018 because of a malfunction in that part of the hadron calorimeter.

\section{Discrimination of signal against backgrounds}
\label{sec:discrimination}
To separate boosted \ttZ and \ttH signal events from background, we use three discriminating variables:
the \pt and \msd of the reconstructed \PZ or Higgs boson candidate AK8 jet, and a global event neural network score.
These variables allow us to define regions that are enriched in \ttZ, regions that are enriched in \ttH, and regions that are dominated by background but kinematically similar to the signal-dominated regions.
The use of the boson candidate \pt allows us to place upper limits on the \ttZ and \ttH cross sections differentially as functions of
$\pt^{\PZ}$ and $\pt^{\PH}$, respectively, and provides additional sensitivity to the effects of BSM physics.

\subsection{Global Event Neural network}

We train a deep neural network (DNN) to discriminate between simulated signal events and simulated \ttjets background events.
The DNN has 50 input variables, two dense hidden layers, and an output layer consisting of three output nodes.
The input variables describe the kinematic properties of reconstructed events and are chosen to provide discrimination power
between \ttZ or \ttH and the \ttjets background, with an emphasis on discrimination against \ttbb events.
All of the input variables are listed in Table~\ref{tab:inputvariables} in Appendix~\ref{app:mva_vars},
and a subset of the input variable distributions, consisting of variables that rank high in the mutual information feature importance metric~\cite{Kreer1957,Shannon1948}, is shown in Fig.~\ref{fig:nn_input_validation}.
These distributions are used to verify that the input variables are well modeled.
Correlations and nonlinear relationships among the input variables are also well understood, which we verify by comparing the expected and observed distributions of the outputs of the final hidden layer of the DNN.

The variables targeting top quark decays include the \PQb tagger scores and jet \pt values,
the invariant masses of sets of two or three jets,
and the spatial separation between pairs of jets or between jets and leptons.
The AK8 jets originating from the decays of boosted \PZ or Higgs bosons are identified using the $\bbbar$ discriminant score,
the number of \PQb{}-tagged AK4 jets overlapping the AK8 jet, and the separations between and invariant masses of the
candidate AK8 jet and various leptons and jets presumed to come from the \ttbar decay products.
Global event variables, including summed \pt, sphericity, and aplanarity~\cite{spher_aplan}, are calculated using the
selected lepton and AK4 jet momenta and the \ptvecmiss.
We combat overtraining of the neural network by employing a batch normalization layer~\cite{batch_norm} and a dropout layer~\cite{JMLR:v15:srivastava14a}, and by evaluating the training with an independent validation data set, which is 60\% of the size of the training data set.

This neural network is a multiclass classifier
with three output nodes, each trained to identify a single class of events:
\ttZ and \ttH signal, \ttbb background, and \ttLF and \ttcc background.
Only well-recon\-struct\-ed \ttZ and \ttH signal events are used in the training, \ie, the
reconstructed \PZ or Higgs boson candidate is matched to both generator-level \PQb quark daughters
of the \PZ or Higgs boson with $\DR(\text{AK8 jet}, \PQb)<0.6$.
The three outputs for each event are scaled so that their sum is one.
The focal loss function~\cite{lin2018focal} is employed to balance the relative importance of the three classes of events during training despite an imbalance in the number of simulated events in each class.

\begin{figure}
  \centering
  \includegraphics[width=\onecolwidth]{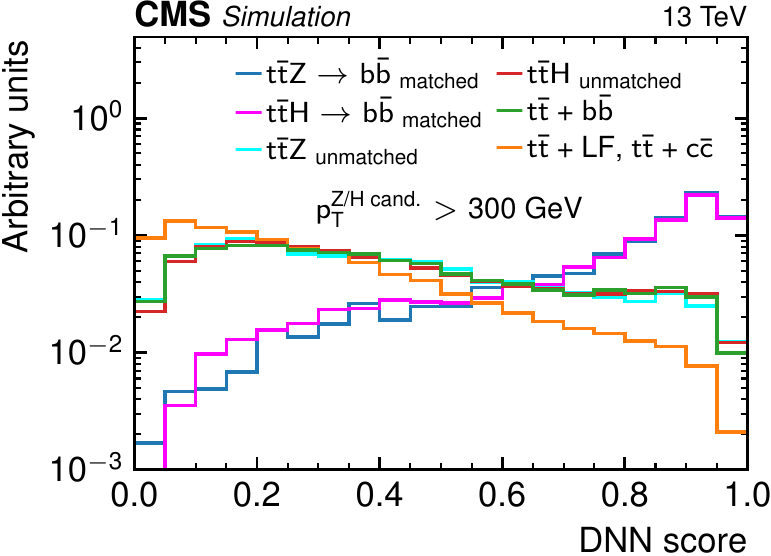}
  \caption{
  The simulated DNN score distributions, normalized to unit area, for well-recon\-struct\-ed \ttZ and \ttH signal events in which the reconstructed \PZ or Higgs boson candidate is matched to both generator-level \PQb quark daughters of the \PZ or Higgs boson;
  the remaining \ttZ and \ttH events;
  background events from \ttbb; and background events from \ttcc and \ttLF. The events satisfy the baseline selection requirements described in Section~\ref{sec:selection} and
  contain a \PZ or Higgs boson candidate with $\pt > 300\GeV$.
  }
\label{fig:NN}
\end{figure}

Although multiple classes are used during training, only the signal-oriented output node is used in the analysis for the purpose of distinguishing background from signal.
Even though they are not directly used, the two background-oriented nodes do affect the signal-oriented output via the scaling, and their presence improves the discrimination power compared to a binary-classification neural network.
Figure~\ref{fig:NN} shows the DNN output score for well-recon\-struct\-ed \ttZ and \ttH signal events, for \ttZ and \ttH events that either do not contain a \PZ or Higgs boson decaying to $\bbbar$ or are not well reconstructed, for \ttbb background events, and for \ttcc and \ttLF background events.
The distributions of well-recon\-struct\-ed signal events show strong separation from those of background events, especially \ttcc and \ttLF.
All \ttZ and \ttH events, even those that either do not contain $\PZ/\PH\to\bbbar$ or are not well reconstructed, are included as signal events in the analysis, even though they are separated in Fig.~\ref{fig:NN}.
Because our sensitivity to BSM effects is dominated by high-\pt events, Fig.~\ref{fig:NN} only includes events with $\pt^{\PZ/\PH} > 300\GeV$.

\subsection{Analysis bins}

Data collected in 2016, 2017, and 2018 are treated separately, and are divided among bins based on the \pt and \msd of the reconstructed \PZ or Higgs boson candidate and the DNN score.
The DNN score provides signal-rich and background-rich regions, enhancing the sensitivity of the analysis.
The \msd provides a region enriched in \ttZ in the vicinity of the \PZ boson mass, a region enriched in \ttH in the vicinity of the Higgs boson mass, and two sidebands with higher and lower \msd that are kinematically similar but depleted in \ttZ and \ttH signal.
The sensitivity to BSM effects is enhanced at high \pt, as shown in Fig.~\ref{fig:analysis:eftimpact}, so binning based on the reconstructed \pt of the boson candidate improves the resulting constraints on the WCs.

Events from each year are divided among three bins based on the \pt of the reconstructed \PZ or Higgs boson candidate: 200--300, 300--450, and ${>}450\GeV$,
motivated by the simplified template cross section binning definition for \ttH~\cite{deFlorian:2016spz}.
Within each \pt range, events are assigned to one of six bins based on the DNN score.
The DNN score binning is unique to each \pt bin in each year, and is chosen so that the \ttZ and \ttH signal events are divided among six intervals bounded by 0, 5, 25, 35, 50, 70, and 100\% of the total expected yield of well-recon\-struct\-ed signal events, ordered from low to high DNN score.

\begin{figure}
  \centering
  \includegraphics[width=0.48\textwidth]{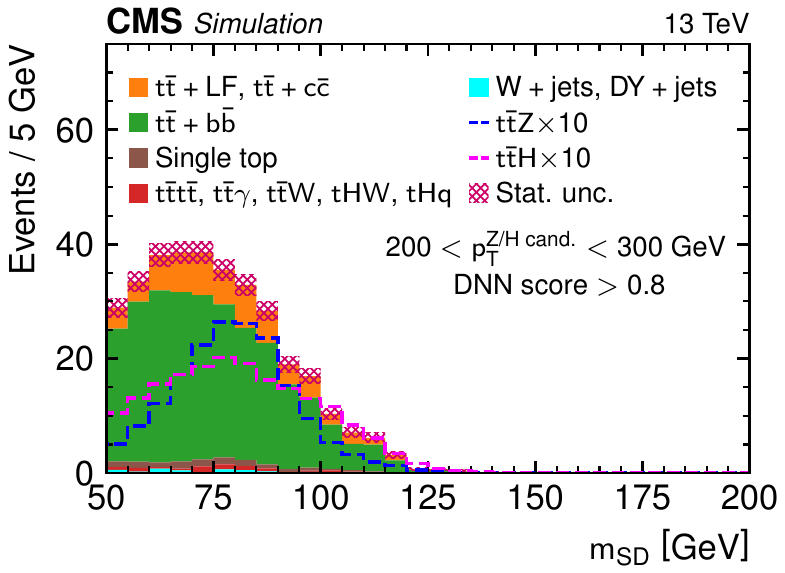}
  \includegraphics[width=0.48\textwidth]{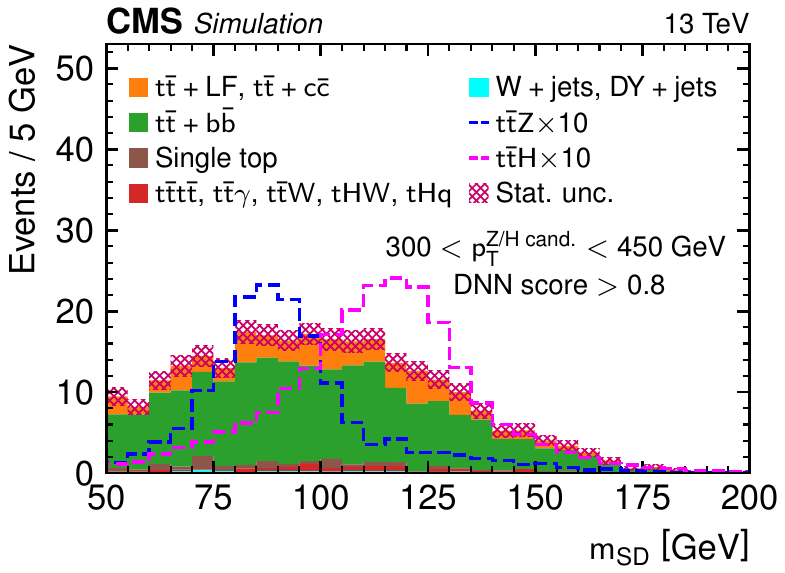}
  \includegraphics[width=0.48\textwidth]{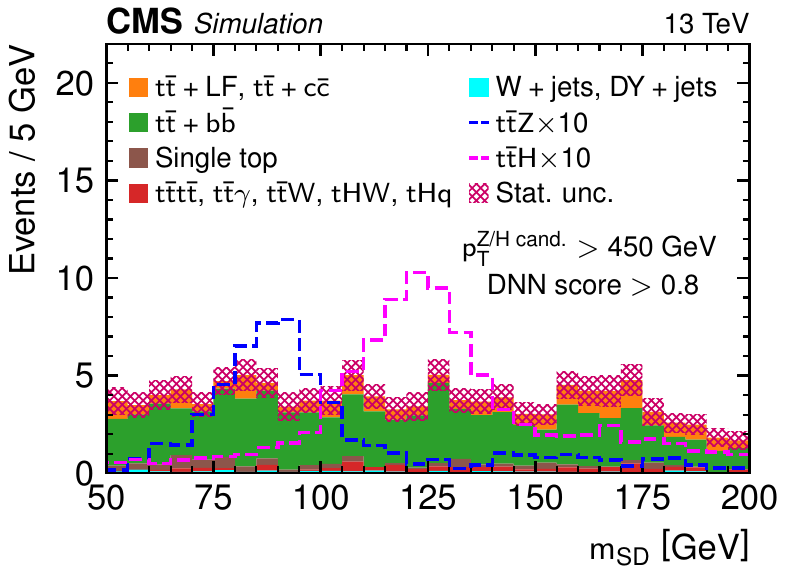}
  \caption{Soft-drop mass distributions from simulation for background (solid histograms) and signal (dashed histograms) of $\PZ/\PH\to\bbbar$ candidate jets in three \pt ranges: 200--300\GeV (\cmsUpperLeft), 300--450\GeV (\cmsUpperRight), and ${>}450\GeV$ (lower) in simulated samples with DNN score ${>} 0.8$.
  The signal distributions represent the SM prediction scaled up by a factor of 10 for easier comparison with the backgrounds.
  The signal distributions include well-recon\-struct\-ed \ttZ and \ttH events as well as \ttZ and \ttH events that either do not contain $\PZ/\PH\to\bbbar$ or are not well reconstructed.
  The red hatched bands correspond to the statistical uncertainty in the background.
  }
\label{fig:msd_zh}
\end{figure}

Within each \pt and DNN score bin, the data are divided among four \msd bins: 50--75, 75--105, 105--145, and 145--200\GeV.
In the lowest \pt bin, the two highest mass bins are merged to $105 < \msd < 200\GeV$ because of low signal and background yields.
Figure~\ref{fig:msd_zh} shows the \msd distributions of reconstructed \PZ or Higgs boson candidates in three \pt ranges for simulated signal and background events.
The $75 < \msd < 105\GeV$ bin provides a \ttZ enriched region for all candidate AK8 jet \pt values,
while the $105 < \msd < 145\GeV$ bin is enriched in \ttH for $\pt > 300\GeV$.
The \ttH events with $\pt < 300\GeV$ do not show a clear peak in \msd around the Higgs boson mass, because its decay products are less likely to be fully contained in the AK8 jet.
Binning using the boson candidate \pt and \msd and the DNN score results in a total of 66 bins for each data-taking year.

\begin{figure}
  \centering
  \includegraphics[width=0.48\textwidth]{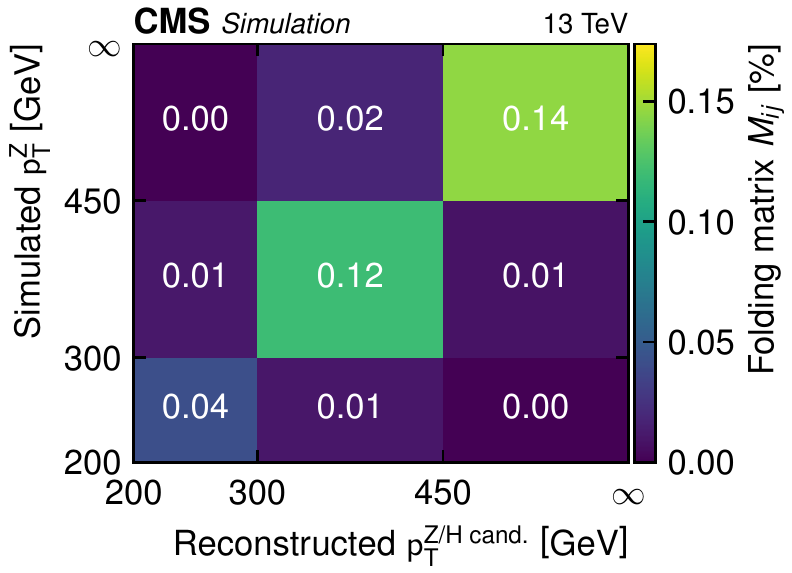}
  \includegraphics[width=0.48\textwidth]{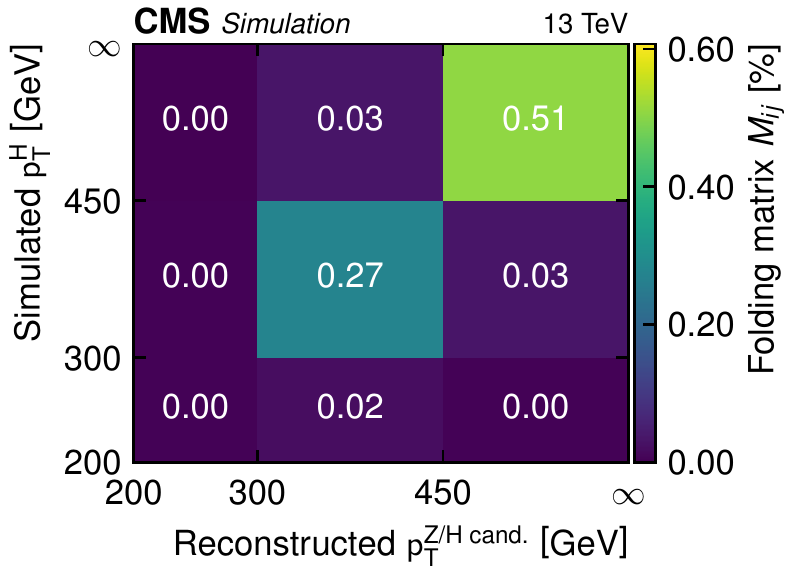}
  \caption{
  The percentage of simulated \ttZ (\cmsLeft) and \ttH (\cmsRight) signal events that satisfy the baseline event selection requirements as well as the DNN and mass requirements
  in bins defined by the reconstructed AK8 jet \pt ($x$ axis) and simulated $\pt^{\PZ/\PH}$ ($y$ axis).
  The rightmost and uppermost bins are unbounded.
  The value of each bin is the ratio of the event yield in the bin to the total number of simulated signal events with a simulated $\pt^{\PZ/\PH}$ in the same $y$-axis bin, including all decay modes of the top quark, \PZ boson, and Higgs boson.
  The color scale to the right shows the meaning of the histogram colors.
  }
\label{fig:diffXSresponse}
\end{figure}

When setting upper limits on the differential cross sections as functions of $\pt^{\PZ}$ and $\pt^{\PH}$, the
\ttZ and \ttH simulated signal samples are each divided into four subsamples according to the \pt of the simulated \PZ or Higgs boson, respectively.
The $\pt^{\PZ}$ and $\pt^{\PH}$ bins are 0--200, 200--300, 300--450, and ${>}450\GeV$.
The three highest \pt bins match the candidate AK8 jet \pt bins, and are used to obtain the differential cross section results.
In events that fall into the generator-level ${\pt^{\PZ/\PH} < 200\GeV}$ bin, the \PZ or Higgs boson does not have sufficient boost to be reconstructed as a single AK8 jet.
Therefore, the rate of those events is not treated as an unconstrained parameter, but rather is fixed to the SM expectation.
For the measurement of the signal strengths of the \ttZ and \ttH processes with generator-level $\pt^{\PZ/\PH}>200\GeV$, the subsample with generator-level $\pt^{\PZ/\PH} < 200\GeV$ is fixed to the SM expectation, but it is unconstrained for the measurement of the WCs.
Figure~\ref{fig:diffXSresponse} shows the expected distribution of the candidate AK8 jet \pt in each of the three highest \pt subsamples of
SM \ttZ and \ttH production for the subset of events that fall into the two or three highest DNN bins and have an AK8 jet mass consistent with \PZ or Higgs boson decay.
These two-dimensional distributions serve as the folding matrices $M_\mathrm{ij}$, which describe the relationship between the simulated and reconstructed distributions, when setting upper limits on the differential cross sections.

\section{Systematic uncertainties}
\label{sec:systematics}
{\tolerance=220
The systematic uncertainties in the measurements
arise from theoretical and experimental sources and may affect both the overall rates of the signals and backgrounds, as well as the shapes of the templates used in the fit to the data.
The systematic uncertainties are treated as nuisance parameters in the final fit to the data.
Except where otherwise noted, the nuisance parameters are given Gaussian constraints in the fit.
The theoretical uncertainties are all treated as fully correlated across the entire data-taking period, while some of the experimental uncertainties are treated independently in each data-taking year and others are fully correlated.
\par}

We account for the statistical uncertainty in the simulation caused by the finite size of the simulated data sets
using the approach described in Refs.~\cite{Barlow-Beeston,Conway:2011in}.

\subsection{Theoretical uncertainties}

The signal and background processes are scaled according to their expected cross sections, which are calculated at NLO or higher order as described in Section~\ref{sec:samples}.
The uncertainties in these cross sections arise from the choice of the renormalization (\muR) and factorization scales (\muF) as well as from uncertainty in the PDFs, including uncertainty in the value of the strong coupling constant \alpS used in the PDFs.
The scales \muR and \muF are varied up and down by a factor of two, and the PDFs and \alpS are varied following the PDF4LHC prescription~\cite{Butterworth:2015oua}.
The resulting variation in the cross sections is treated as a set of nuisance parameters with log-normal constraints depending on the process and the initial state.
These nuisance parameters only affect the overall yield of each process.

The choice of \muR and \muF scales, as well as the PDF, may also affect the simulation of the kinematic distributions of the particles in the final state.
In the simulation of the matrix elements, the scales are varied up and down by a factor of two, and the PDF is varied within its uncertainties~\cite{Butterworth:2015oua}.
Similarly, the scales that control the amount of initial- and final-state radiation are varied up and down independently by a factor of two.
Although the parton shower scale variations affect all processes, special handling is required for the \ttbb background.
Variations in the parton shower scales affect the fraction of \ttbb events within the \ttjets background.
The effect on the \ttbb yield is calculated within the \ttjets (5FS) MC sample and then propagated to the \ttbb (4FS) MC sample.
Because the effect of these uncertainties on the overall yields has already been accounted for, these variations only affect the acceptance or the shape of the distributions in the inclusive phase space.
This is treated as a set of nuisance parameters, depending again on the process and the initial state.

For the \POWHEG samples, the matching between the matrix element and parton shower simulation of the \ttjets backgrounds is governed by the \hdamp parameter.
The \ttjets backgrounds are simulated with ${\hdamp = 1.379\mtop}$, where \mtop is the mass of the top quark, and smaller simulated samples are produced with ${\hdamp = 0.874\mtop}$ and ${\hdamp = 2.305\mtop}$ to evaluate the effects of varying this parameter~\cite{Sirunyan:2019dfx}.
We treat this uncertainty as a nuisance parameter with a log-normal constraint that affects the overall yields of the \ttjets backgrounds.

Any remnants of the $\Pp\Pp$ collision that do not participate in the hard-scattering process are referred to as the ``underlying event''.
The simulation is tuned to reproduce the kinematic distributions of the underlying event~\cite{Sirunyan:2019dfx}.
The effects of the residual uncertainty in the tune are evaluated in separate simulations of the \ttjets backgrounds.
The uncertainty in the underlying-event tune is treated as a nuisance parameter with a log-normal constraint affecting the overall yields of the \ttjets backgrounds.

The cross section of the \ttbb background is treated as an unconstrained nuisance parameter.
Moreover, the rate of the \tttwob subset of the \ttbb background is varied by up to 50\% because of the uncertainty in modeling the gluon fragmentation process~\cite{ttbb:2020_frag_hf}.

The \ttcc production cross section has been measured with an uncertainty of approximately 20\%~\cite{ttcc:top20003}.
However, we assume a larger uncertainty of 50\% because of the differences between the phase space studied in Ref.~\cite{ttcc:top20003} and the phase space relevant for this study.

\subsection{Experimental uncertainties}

The templates in the fit are scaled to their expected yields using the integrated luminosity represented by our data set.
The integrated luminosities recorded in 2016, 2017, and 2018 are measured with uncertainties of 1.2, 2.3, and 2.5\%, respectively~\cite{CMS-LUM-17-003,CMS-PAS-LUM-17-004,CMS-PAS-LUM-18-002}.
These measurements are mostly uncorrelated from year to year, and we treat the uncertainty in the integrated luminosity with a set of nuisance parameters with log-normal constraints representing the correlated and uncorrelated components.
This does not affect the shapes of the templates, while all of the other experimental uncertainties do affect the template shapes.

The distribution of the number of pileup interactions in simulated events is reweighted to match the observed pileup distribution in data,
assuming a total inelastic $\Pp\Pp$ cross section of 69.2\unit{mb}~\cite{Sirunyan:2018nqx}.
This cross section is varied up and down by 4.6\%, and the resulting effects on the simulation are treated as independent nuisance parameters for each year.
This accounts for year-to-year variations in the differences between the interaction vertex multiplicity distribution in data and simulation.

The simulation is corrected, using \pt- and $\eta$-dependent scale factors,
for residual differences between data and simulation in the reconstruction, identification, and isolation of the lepton, and in the trigger efficiency.
The uncertainties in these scale factors are treated as independent nuisance parameters for each year.
The trigger efficiency scale factor includes a correction and accompanying uncertainty component caused by a gradual drift in the timing of the inputs to the ECAL level-1 trigger in the forward region during 2016 and 2017~\cite{CMS:2020cmk}.

The jet energy scales for AK4 and AK8 jets in simulation are corrected to match data, and these corrections are varied within their uncertainties~\cite{CMS_JES}.
These effects are split among 11 independent subsources of uncertainty, some of which are correlated among the years.
The uncertainty subsources affecting both AK4 and AK8 jets are treated as correlated.
Similarly, the energy resolutions of AK4 and AK8 jets are known to be worse in data than in simulation, so the simulated jet energies are smeared to improve their agreement with data~\cite{CMS_JES}.
The amount of smearing is varied within the uncertainty in the required amount of smearing, and the effects of this variation are handled independently for each year.
Additionally, the smearing procedure is performed independently for AK4 and for AK8 jets, so the corresponding uncertainties are uncorrelated.
Corrections are also applied to the scale and resolution of \msd for AK8 jets, and these corrections are varied within their uncertainties.
Corrections to the \msd values from simulation are derived from a comparison of simulated and measured samples of events in a region enriched with merged $\PW\to\PQq\PAQq^\prime$ decays from \ttbar events~\cite{CMS-PAS-JME-16-003}.
The \msd corrections remove a residual dependence on the jet \pt, and match the simulated jet mass scale and resolution to those observed in data.
These corrections are varied within their uncertainties. 
The effects of these variations are treated as independent nuisance parameters for each year.

The performance of the \textsc{DeepCSV} \PQb tagger is slightly different in simulation than in data.
We correct the simulation separately for the tagging of \PQb jets, the tagging of \PQc jets, and the misidentification of light-flavor jets using efficiency scale factors~\cite{DeepCSV}.
Uncertainties in these scale factors arise from a variety of effects, including variations in the jet energy scales, the finite size of the samples used to measure the scale factors, and background contamination in the data regions used to measure the scale factors.
The uncertainties in the scale factors for each effect are treated as nuisance parameters, some of which are correlated from year to year and some of which are independent for each year.
Similarly, the efficiency of the \textsc{DeepAK8} $\bbbar$ tagger is also corrected in simulation to better match the data.
The scale factors are obtained using a collection of AK8 jets enriched in gluon splitting to a \bbbar pair based on the presence of secondary vertices~\cite{JME-18-002,DeepCSV}.
The uncertainty in this correction is treated independently for each year.

\section{Results}
\label{sec:results}
Using the discriminating variables described in Section~\ref{sec:discrimination}, namely
the boson candidate AK8 jet \pt, its \msd, and the full event neural network score,
we measure the signal strengths of boosted \ttZ and \ttH production,
determine 95\% confidence level (\CL) upper limits on \ttZ and \ttH cross sections,
and constrain the effects of BSM physics arising from dimension-six operators in an EFT framework.

To perform these measurements, we use binned maximum likelihood fits, in which the signal and background model is compared to the data in each bin.
The likelihood function used to fit the data is the Poisson likelihood with a rate described by the signal and background model,
\ifthenelse{\boolean{cms@external}}
{ 
	\begin{multline*}
		L_\text{data} \left(\boldsymbol\theta_\text{POI}, \boldsymbol\theta_\text{nuis} \mid D_\mathrm{i}\right) = {} \\ \prod_\mathrm{i} \frac{\lambda_\mathrm{i} \left(\boldsymbol\theta_\text{POI}, \boldsymbol\theta_\text{nuis}\right)^{D_\mathrm{i}} e^{-\lambda_\mathrm{i} \left(\boldsymbol\theta_\text{POI}, \boldsymbol\theta_\text{nuis}\right)}}{D_\mathrm{i} !},
	\end{multline*}
} 
{ 
	\begin{equation*}
		L_\text{data} \left(\boldsymbol\theta_\text{POI}, \boldsymbol\theta_\text{nuis} \mid D_\mathrm{i}\right) = \prod_\mathrm{i} \frac{\lambda_\mathrm{i} \left(\boldsymbol\theta_\text{POI}, \boldsymbol\theta_\text{nuis}\right)^{D_\mathrm{i}} e^{-\lambda_\mathrm{i} \left(\boldsymbol\theta_\text{POI}, \boldsymbol\theta_\text{nuis}\right)}}{D_\mathrm{i} !},
	\end{equation*}
} 
where
\begin{equation*}
	\lambda_\mathrm{i} \left(\boldsymbol\theta_\text{POI}, \boldsymbol\theta_\text{nuis}\right) = S_\mathrm{i} \left(\boldsymbol\theta_\text{POI}, \boldsymbol\theta_\text{nuis}\right) + B_\mathrm{i} \left(\boldsymbol\theta_\text{nuis}\right)
\end{equation*}
and where $\mathrm{i}$ is a bin index, which runs over all the bins, $D_\mathrm{i}$ is the number of observed events, $S_\mathrm{i}$ is the number of events predicted by the signal model, $B_\mathrm{i}$ is the number of predicted background events, $\boldsymbol\theta_\text{POI}$ is the set of parameters of interest (POIs), and $\boldsymbol\theta_\text{nuis}$ is the set of nuisance parameters.
The background model depends only on the nuisance parameters, while the signal model depends on both the nuisance parameters and the POIs.
The overall likelihood function is the product of $L_\text{data}$ and the constraint factors for each of the nuisance parameters, which introduce a likelihood penalty when the nuisance parameters shift from their nominal values.
The constraint factors are Gaussian with a few exceptions, which are detailed in Section~\ref{sec:systematics}.
The \ttbb rate is only constrained to be positive and otherwise is unconstrained in the fit.
The factors representing the statistical uncertainties in the simulation use the approach described in Ref.~\cite{Barlow-Beeston}.

For each fit, the likelihood function is maximized in terms of both the POIs and the nuisance parameters in order to obtain the best fit values of the POIs.
Confidence intervals for the POIs are obtained by profiling the likelihood ratio~\cite{CMS-NOTE-2011-005}, where the denominator is the maximized likelihood and the numerator is the maximized likelihood given a fixed value for one or more POIs.
We perform one-dimensional and two-dimensional scans of the likelihood ratio in which one or two POIs are repeatedly fixed to specific values while the likelihood is maximized as a function of all the remaining parameters.
The 95\% \CL upper limits are set with the asymptotic approximation~\cite{asymptotic_limits} of the modified frequentist \CLs criterion~\cite{JunkCLs:1999,ReadCls:2002},
in which the profile likelihood ratio is modified for upper limits~\cite{CMS-NOTE-2011-005} and used as the test statistic.

We validate the results by comparing them to the results we would obtain by considering each data-taking year independently.
The results from each year are in agreement with each other and with the full results.

\begin{figure*}
  \centering
  \includegraphics[width=\textwidth]{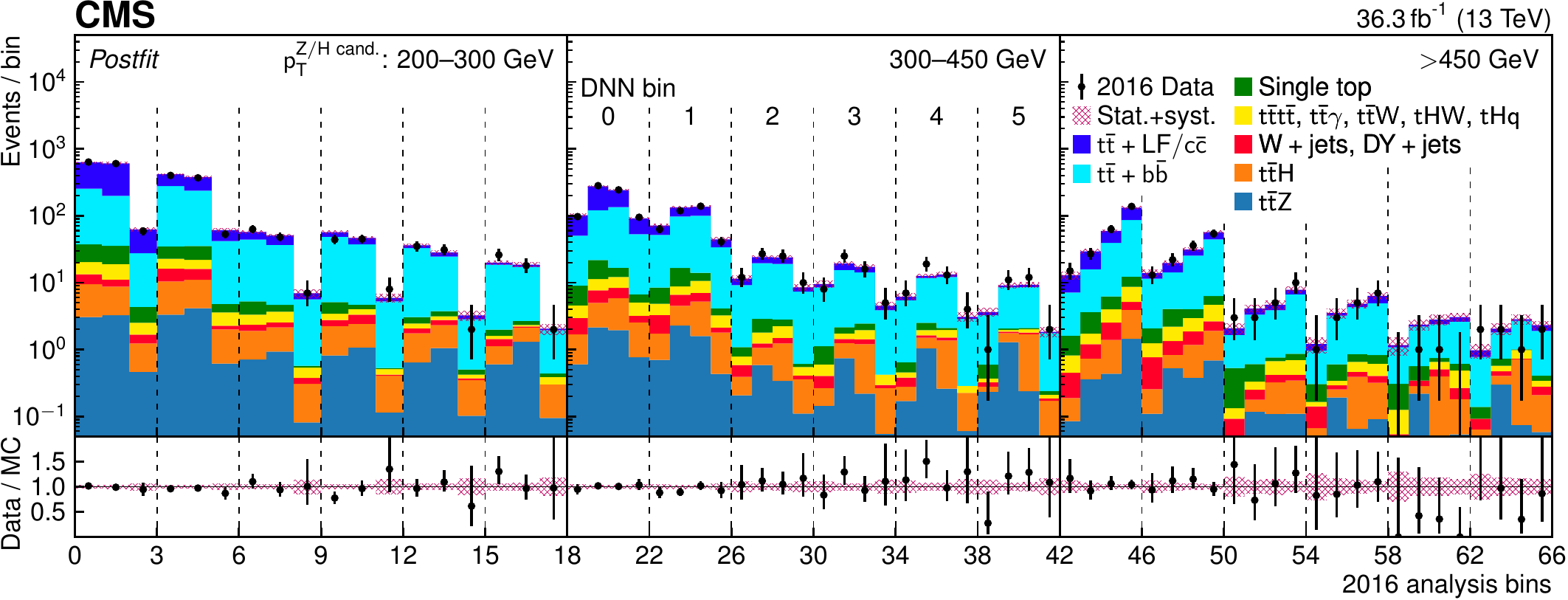}
  \includegraphics[width=\textwidth]{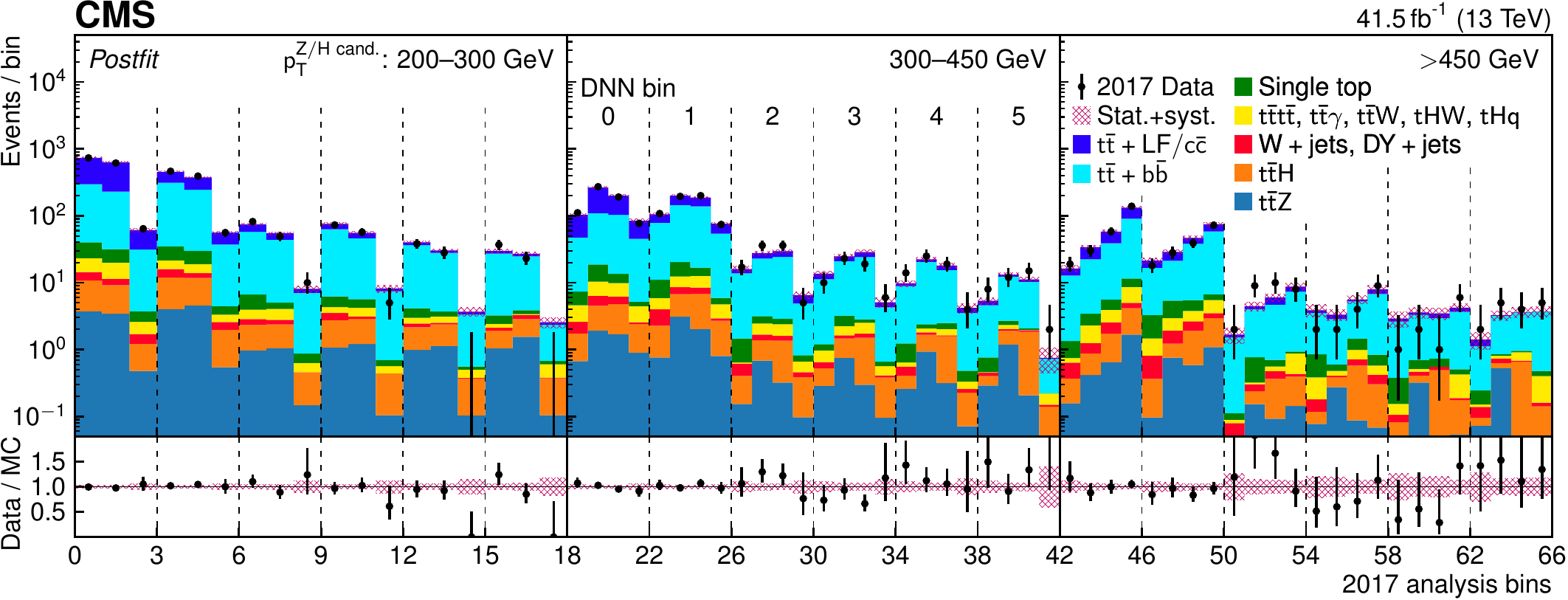}
  \includegraphics[width=\textwidth]{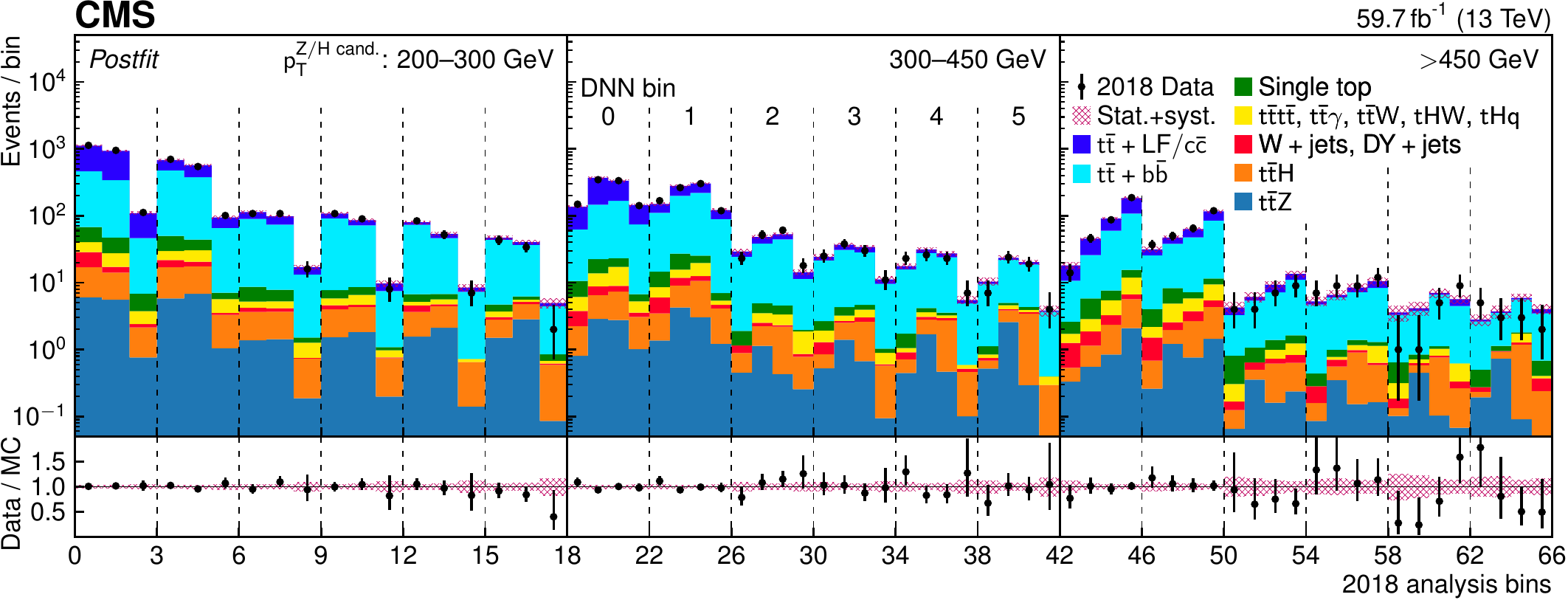}
  \caption{Postfit expected (solid histograms) and observed (points) yields for the 2016 (upper), 2017 (middle), and 2018 (lower) data-taking periods in each analysis bin.
  In the fit, the \ttZ and \ttH signal cross sections are fixed to the SM predictions.
  The analysis bins are defined as functions of the DNN score, and the \pt and \msd of the boson candidate AK8 jet.
  The largest three groupings of bins in each year are defined by the AK8 jet \pt.
  The next six groups are defined by the DNN score, and the smallest groups of three or four bins with the same \pt and DNN score correspond to the AK8 jet \msd.
  The vertical bars on the points show the statistical uncertainty in the data.
  The lower panels in each plot give the ratio of the data to the sum of the MC predictions, with the red band representing the total uncertainty in the MC prediction.
  }
\label{fig:fit_data}
\end{figure*}

The observed yields in the 198 analysis bins are shown in Fig.~\ref{fig:fit_data}, together with the expected yields
from signal and background processes as predicted by the SM simulation, after fitting to the data (postfit).

\subsection{Signal strengths and upper limits on differential cross sections}
\label{subsec:XS}

\begin{figure}
  \centering
  \includegraphics[width=\onecolwidth]{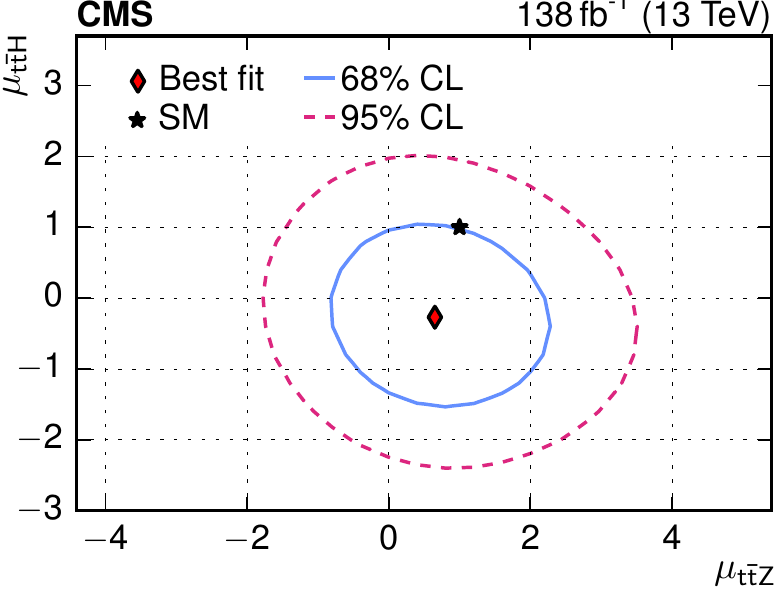}
  \caption{The observed best fit values (diamond) and SM predictions (star) for the signal strength modifiers $\mu_{\ttH}$ versus $\mu_{\ttZ}$ for generator-level $\pt^{\PH}$ or $\pt^{\PZ} > 200\GeV$.  The solid blue and dashed red contours show the 68 and 95\% \CL regions, respectively.}
\label{fig:signalStrengths}
\end{figure}

To facilitate comparisons with theoretical predictions, we isolate and remove the effects of limited detector acceptance and response to \ttZ and \ttH production using a maximum likelihood unfolding technique as described in Section~5 of Ref.~\cite{Sirunyan:2018sgc}.
Using this technique, we perform a measurement of the signal strengths of boosted \ttZ and \ttH production, \ie, the ratios of the measured \ttZ and \ttH production cross sections to the cross sections predicted by the SM as defined in Section~\ref{sec:samples}.
The measurement is performed in the region of generator-level phase space defined by the rapidity requirement $\abs{y_{\PZ/\PH}} < 2.5$
and $\pt^{\PZ/\PH} > 200\GeV$.
The POIs that affect the signal model are the \ttZ and \ttH production signal strengths, $\mu_{\ttZ}$ and $\mu_{\ttH}$, respectively.
The results are shown in Fig.~\ref{fig:signalStrengths} and Table~\ref{tab:results:smsum}, and
the impacts $\Delta{\mu}$ on $\mu_{\ttZ}$ versus $\mu_{\ttH}$ from the major sources of systematic uncertainty are listed in Table~\ref{tab:signal_uncertainty}.
The $\Delta{\mu}$ for \ttZ and \ttH are obtained by freezing one group of nuisance parameters at a time, and subtracting, in quadrature, the resulting 68\% interval from the full fit interval.
The observed correlation between $\mu_{\ttZ}$ and $\mu_{\ttH}$ is $-10\%$.
The observed yield for the \ttbb background is higher than the nominal SM prediction from the MC simulation, but it is
consistent with the findings of other CMS measurements~\cite{ttbb:top18002,ttbb:top18011,ttcc:top20003}.

\begin{table*}
\topcaption{
        The observed and expected best fit ($\pm 1$ standard deviation) signal strength modifiers $\mu_{\ttZ}$ and $\mu_{\ttH}$ for simulated $\pt^{\PZ}$ or $\pt^{\PH} > 200\GeV$.
        The observed uncertainties are broken down into the components arising from the limited size of the data, the limited size of the simulation samples,
        experimental uncertainties, and theoretical uncertainties.
}
\label{tab:results:smsum}
\centering
\renewcommand{\arraystretch}{1.5}
	\begin{scotch}{lcccccc}
	Signal strength & Observed & Stat. & MC stat. & Exp. syst. & Theo. syst. & Expected \\
  \hline \\[-3.5ex]
	$\mu_{\ttZ}$  & ${0.65\,}_{-0.98}^{+1.04}$  & ${}_{-0.75}^{+0.80}$ & ${}_{-0.38}^{+0.36}$ & ${}_{-0.31}^{+0.38}$ & ${}_{-0.38}^{+0.43}$  & ${1.00\,}_{-0.84}^{+0.91}$\\
	$\mu_{\ttH}$  & ${-0.27\,}_{-0.83}^{+0.86}$  & ${}_{-0.65}^{+0.72}$ & ${}_{-0.33}^{+0.31}$ & ${}_{-0.19}^{+0.19}$ & ${}_{-0.35}^{+0.28}$ & ${1.00\,}_{-0.72}^{+0.79}$\\
\end{scotch}
\end{table*}

\begin{table}
	\topcaption{
		The magnitudes of the major sources of systematic uncertainty in the measurement of the signal strength modifiers $\mu_{\ttZ}$ and $\mu_{\ttH}$ for simulated $\pt^{\PZ}$ or $\pt^{\PH} > 200\GeV$.
	}
	\label{tab:signal_uncertainty}
	\centering
	\renewcommand{\arraystretch}{1.5}
	\begin{scotch}{lcc}
		Source of uncertainty           & $\Delta{\mu}_{\ttZ}$  & $\Delta{\mu}_{\ttH}$ \\
		\hline
		\ttcc cross section              & ${}_{-0.23}^{+0.27}$  & ${}_{-0.13}^{+0.14}$   \\
		\ttbb cross section              & ${}_{-0.24}^{+0.18}$  & ${}_{-0.22}^{+0.16}$   \\
		\tttwob cross section            & ${}_{-0.03}^{+0.03}$  & ${\pm}0.09$            \\
		\muR and \muF scales             & ${}_{-0.14}^{+0.12}$  & ${}_{-0.15}^{+0.11}$   \\
		Parton shower                    & ${}_{-0.17}^{+0.16}$  & ${}_{-0.06}^{+0.07}$   \\
		\PQb tagging efficiency          & ${}_{-0.13}^{+0.25}$  & ${\pm}0.10$            \\
		$\bbbar$ tagging efficiency      & ${}_{-0.13}^{+0.18}$  & ${}_{-0.04}^{+0.07}$   \\
		Jet energy scale and resolution  & ${\pm}0.11$           & ${}_{-0.12}^{+0.11}$   \\
		Jet mass scale and resolution    & ${\pm}0.10$           & ${\pm}0.08$            \\
	\end{scotch}
\end{table}

Because the uncertainties in the boosted \ttZ and \ttH cross sections are large, we are not able to provide two-sided confidence intervals in individual bins of $\pt^{\PZ}$ and $\pt^{\PH}$.
Instead, we provide 95\% \CL upper limits on the differential cross sections for production of \ttZ and \ttH as functions of $\pt^{\PZ}$ and $\pt^{\PH}$.
We account for smearing of the reconstructed $\pt^{\PZ}$ and $\pt^{\PH}$ compared to the simulated quantities using folding matrices, which are similar to those shown in Fig.~\ref{fig:diffXSresponse}, but expanded to include the DNN score and \msd.
Like the signal strength measurement, these upper limits are established in the region of generator-level phase space defined by $\abs{y_{\PZ/\PH}} < 2.5$
and $\pt^{\PZ/\PH} > 200\GeV$.
Simulated \ttZ and \ttH events are divided into subsamples according to the simulated $\pt^{\PZ}$ and $\pt^{\PH}$, which,
along with the cross section corresponding to each subsample, form the signal model and its POIs.
When setting the upper limit on one differential cross section, the other differential cross sections are profiled.
The 95\% \CL upper limits on the differential cross sections are given in Fig.~\ref{fig:diffXS} and Table~\ref{tab:results:sm_upperlimits}.

\ifthenelse{\boolean{cms@external}}{\begin{figure*}[!p]}{\begin{figure*}}
  \centering
  \includegraphics[width=0.49\textwidth]{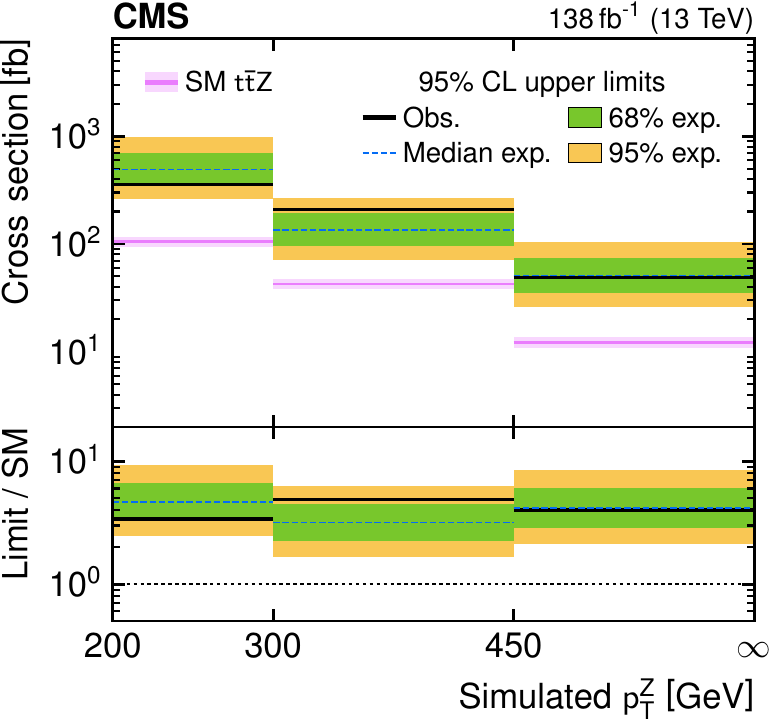}
  \includegraphics[width=0.49\textwidth]{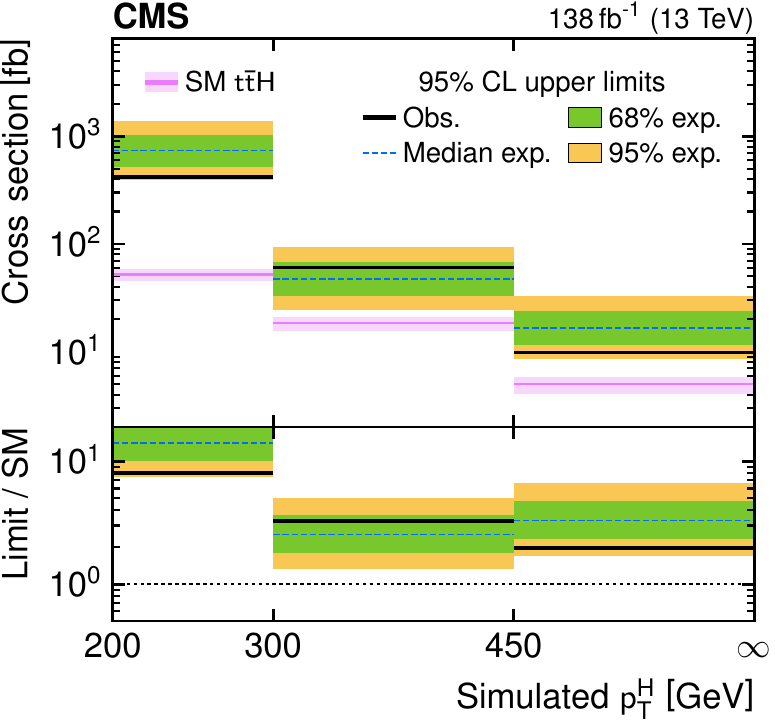}
  \caption{
  Observed and expected 95\% \CL upper limits on the \ttZ (left) and \ttH (right) differential cross sections as a function of the simulated \PZ and Higgs boson \pt.
  The green and yellow bands indicate the regions containing 68 and 95\%, respectively, of the distribution of limits under the SM hypothesis.
  The black lines represent the observed 95\% \CL upper limits.
  The magenta band shows the SM predicted differential cross sections with the $\text{PDF}+\alpS$ and scale uncertainties.
  The lower panel shows the ratio of the observed and expected upper limits on the differential cross sections to the SM differential cross sections.
  The last bin is unbounded.
  }
\label{fig:diffXS}
\end{figure*}

\ifthenelse{\boolean{cms@external}}{\begin{table*}[!p]}{\begin{table*}}
\topcaption{
		Observed and median expected 95\% \CL upper limits on the \ttZ and \ttH differential cross sections and on their ratios to the SM predictions for three $\pt^{\PZ/\PH}$ intervals.
		The range given with the median expected 95\% \CL upper limits indicates the range in which 68\% of the upper limits are expected to fall, assuming the SM hypothesis.
}
\label{tab:results:sm_upperlimits}
\centering
\renewcommand{\arraystretch}{1.5}
\begin{scotch}{ l l c c c c}
    Signal &  $\pt^{\PZ/\PH}$ interval [{\GeVns}] & \multicolumn{2}{c}{95\% \CL upper limit [fb]} & \multicolumn{2}{c}{95\% \CL upper limit / SM} \\
           &                                      & Observed & Expected                           & Observed & Expected                           \\
    \hline
    \ttZ   & 200--300                             & 360      & ${490\,} _{-140}^{+210}$           & 3.4      & ${4.7\,} _{-1.4}^{+2.0}$           \\
           & 300--450                             & 209      & ${135\,} _{-40} ^{+58}$            & 4.9      & ${3.2\,} _{-0.9}^{+1.4}$           \\
           & ${>}450$                             & 49       & ${51\,}  _{-15} ^{+23}$            & 4.0      & ${4.1\,} _{-1.3}^{+1.9}$           \\[\cmsTabSkip]
    \ttH   & 200--300                             & 420      & ${740\,} _{-210}^{+300}$           & 8.0      & ${14.1\,}_{-4.1}^{+5.7}$           \\
           & 300--450                             & 60       & ${47\,}  _{-14} ^{+20}$            & 3.2      & ${2.5\,} _{-0.8}^{+1.1}$           \\
           & ${>}450$                             & 9.8      & ${16.5\,}_{-4.9}^{+7.3}$           & 2.0      & ${3.3\,} _{-1.0}^{+1.5}$           \\
    \end{scotch}
\end{table*}

\subsection{Effective field theory constraints}
\label{subsec:EFTconstraints}

We constrain eight WCs from the LO dimension-six EFT model described in Section~\ref{sec:eft_samples}.
Again, we impose a generator-level requirement on the rapidity of the \PZ or Higgs boson, $\abs{y_{\PZ/\PH}} < 2.5$.
The effects of varying the WCs are computed as a weight $w_\text{EFT}$ for each simulated event in the
\ttZ, \ttH, and \ttbb samples, detailed in Eq.~\eqref{eq:quad_wgt_eqn}.
This weight affects the predicted yields in each analysis bin, as described in Section~\ref{subsec:XS}.
For each bin, we compute the ratio of the expected rate in the EFT model to the expected rate in the LO SM, as a function of the WCs.
The LO SM expected yields are multiplied by this ratio in each bin and by an NLO $K$ factor, which is the ratio of the expected NLO SM yield to the expected LO SM yield.
This produces predicted NLO histograms parameterized in the WCs.
Thus, the signal yield in each bin is expressed as
\ifthenelse{\boolean{cms@external}}
{ 
	\begin{equation*}
		\begin{aligned}
			\text{bin\;yield} &= \text{LO\;SM\;yield}  \times {}\\ &\phantom{{}={}} \frac{\text{NLO\;SM\;yield}}{\text{LO\;SM\;yield}} \times \frac{\text{LO\;EFT\;yield}}{\text{LO\;SM\;yield}}.
		\end{aligned}
	\end{equation*}
} 
{ 
	\begin{equation*}
		\text{bin\;yield} = \text{LO\;SM\;yield}  \times \frac{\text{NLO\;SM\;yield}}{\text{LO\;SM\;yield}} \times \frac{\text{LO\;EFT\;yield}}{\text{LO\;SM\;yield}}.
	\end{equation*}
} 
We maximize the likelihood function in terms of the WCs and the various nuisance parameters representing the systematic uncertainties.

We perform a likelihood scan for each WC while fixing the seven other WCs to their SM value of zero.
With the WCs fixed, we find the maximum likelihood as a function of the nuisance parameters.
We extract 68 and 95\% \CL intervals from the resulting profiled likelihood as shown in Fig.~\ref{fig:EFTresults2}.
We also perform a second scan for each WC in which the seven other WCs are not fixed, so that the likelihood is maximized as a function of the nuisance parameters and seven of the eight WCs.
Both sets of intervals are summarized in Fig.~\ref{fig:eft_results}, and the 95\% \CL intervals are summarized in Table~\ref{tab:EFTresults}.
Because of partial degeneracies in the effects of the WCs, the constraints on the WCs from the two different likelihood scans tend to differ for \cpQM, \cpQa, \cpt, \ctW, and \ctZ.
To further explore this, two-dimensional likelihood scans are performed for pairs of WCs while the remaining six WCs are fixed to zero.
Figure~\ref{fig:EFT2dscans} shows the 68, 95, and 99.7\% \CL regions extracted from the scans for the three pairs of WCs that exhibit the largest postfit correlations.
As shown in Table~\ref{tab:eftOperators}, two of these WC pairs (\cpQa, \cpQM) and (\ctW, \ctZ) apply to the same operators \Opq and \Oub, respectively,
and the other WC pair (\cpt, \cpQM) corresponds to the operators \Opu and \Opq, which have a similar structure, leading to the observed correlations.

\begin{figure*}
  \centering
    \includegraphics[width=0.49\textwidth]{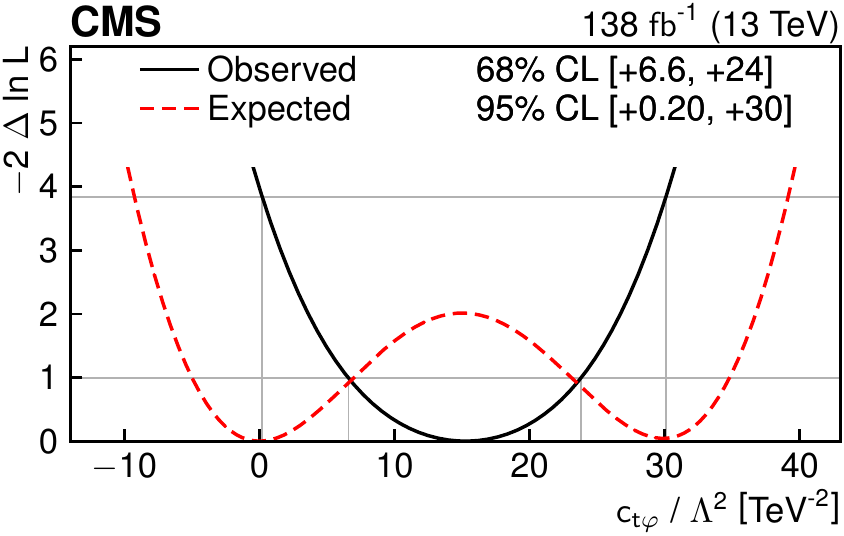}
    \includegraphics[width=0.49\textwidth]{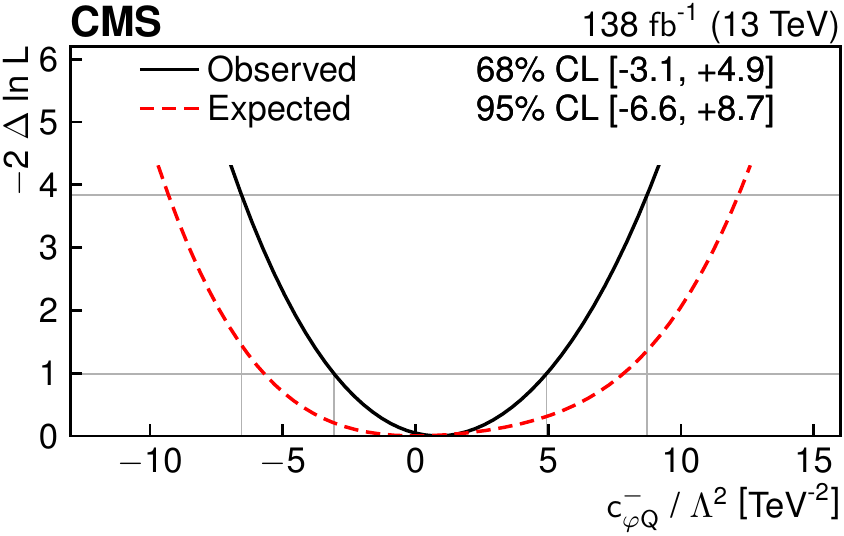}
    \includegraphics[width=0.49\textwidth]{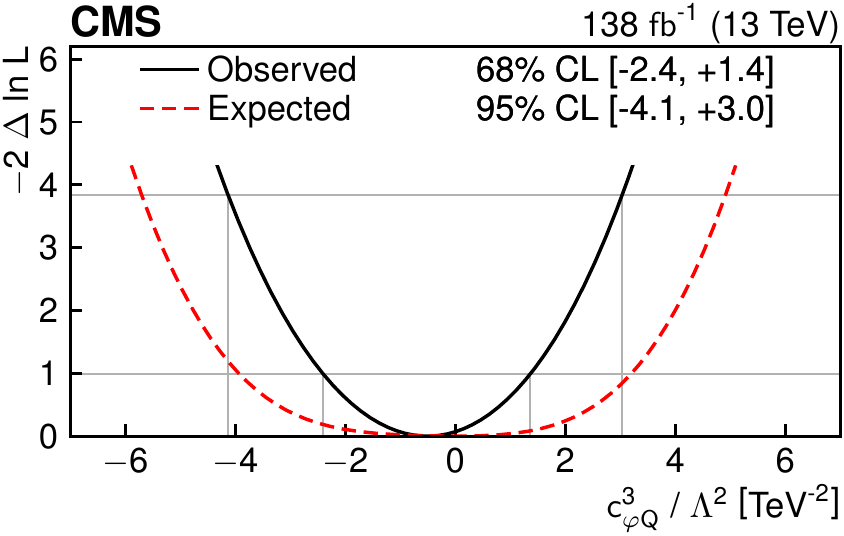}
    \includegraphics[width=0.49\textwidth]{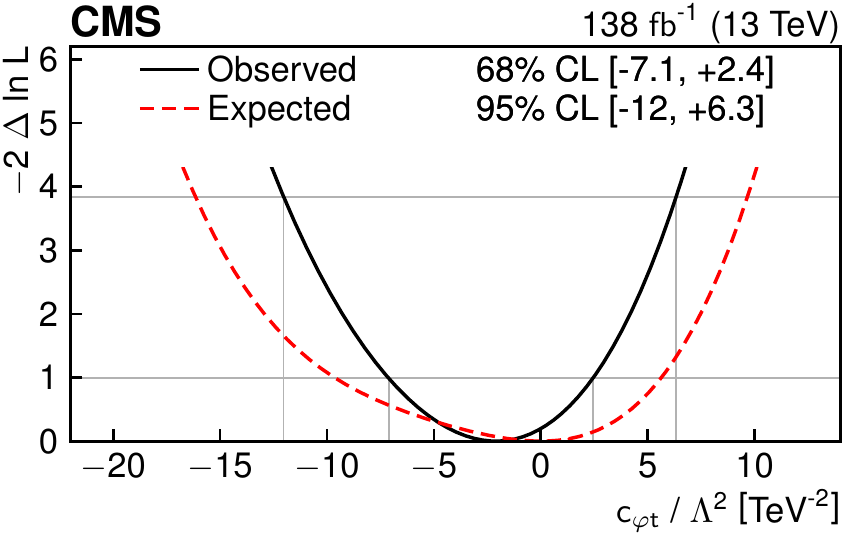}
    \includegraphics[width=0.49\textwidth]{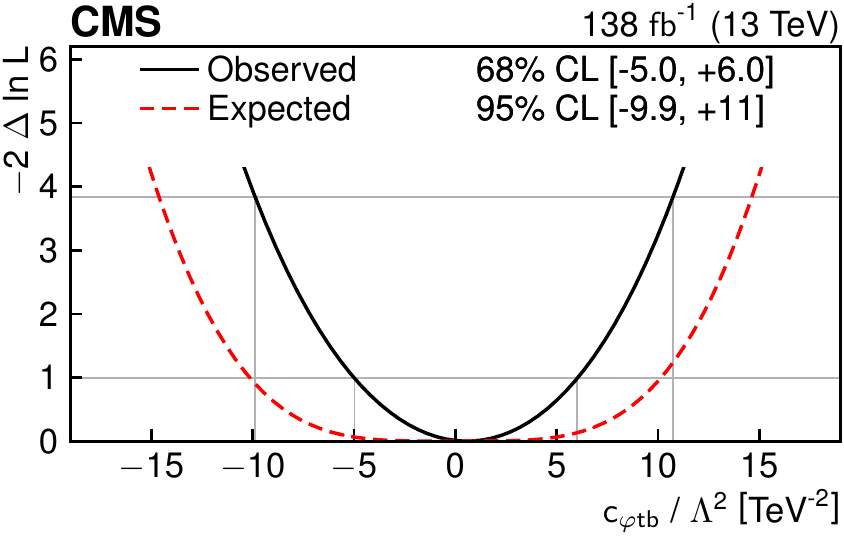}
    \includegraphics[width=0.49\textwidth]{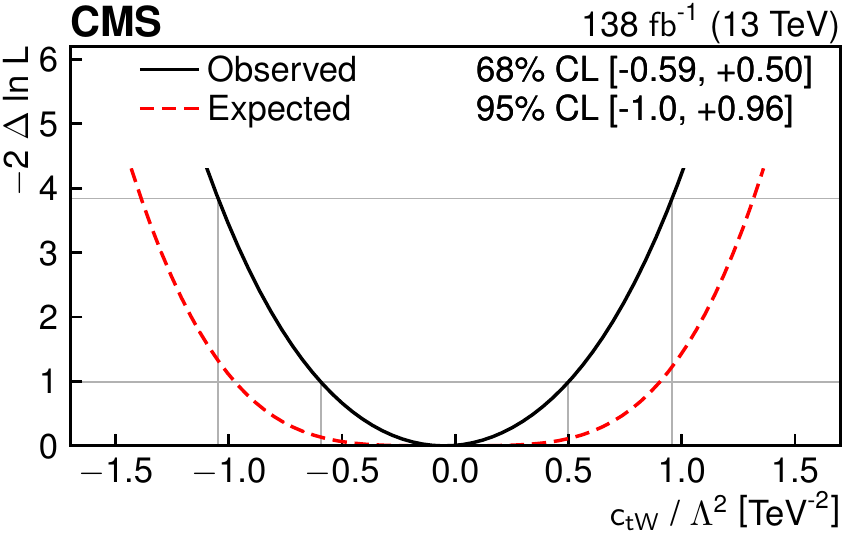}
    \includegraphics[width=0.49\textwidth]{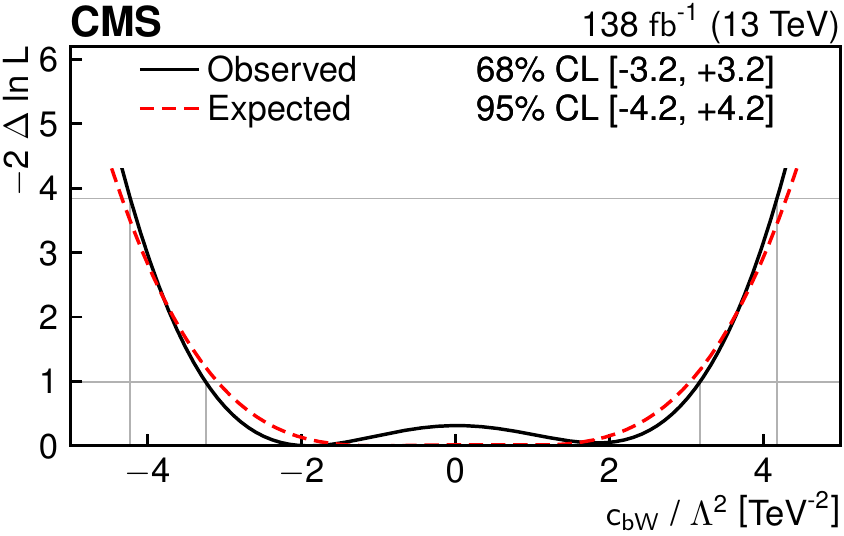}
    \includegraphics[width=0.49\textwidth]{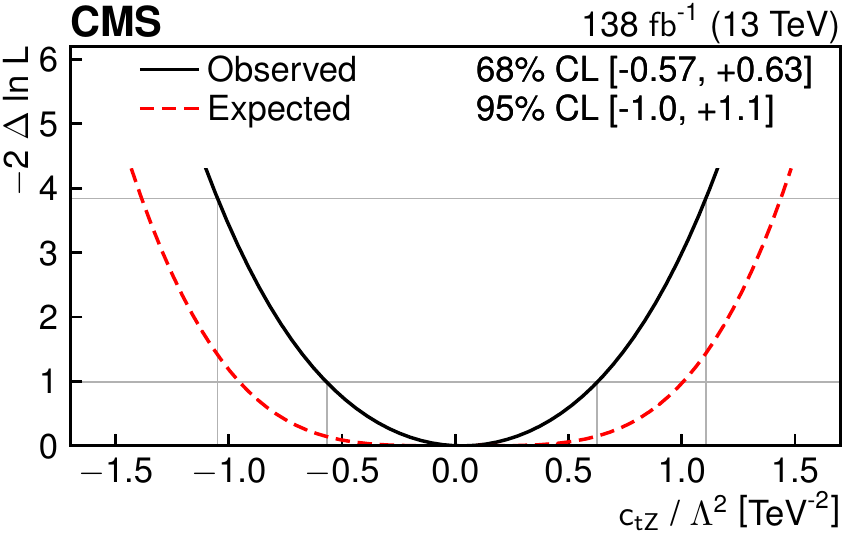}
    \caption{
    Observed (solid black) and expected (dotted red) scans of the negative log-likelihood as a function of each of the eight WCs when the seven other WCs are fixed to their SM values.
    The 68 and 95\% \CL intervals are indicated by thin gray lines.
    }
   \label{fig:EFTresults2}
\end{figure*}

\begin{figure}
  \centering
  \includegraphics[width=\onecolwidth]{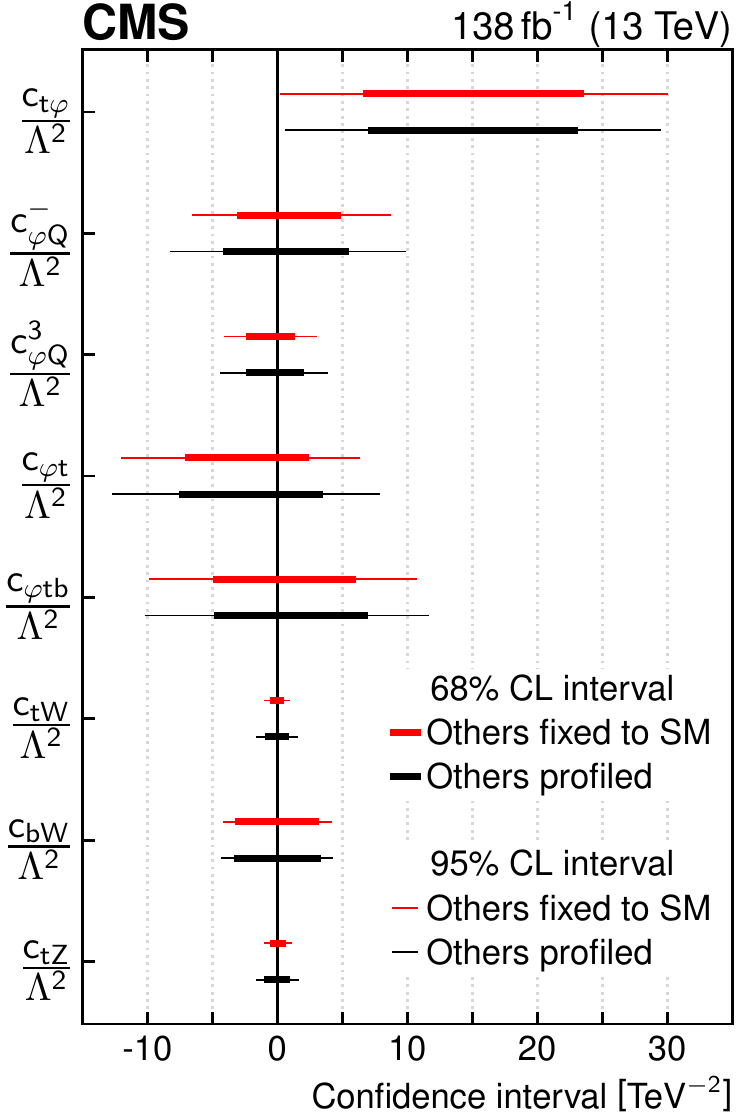}
  \caption{
	The observed 68 and 95\% \CL intervals for the WCs are shown by the thick and thin bars, respectively. The intervals are found by scanning over a single WC while either profiling the other seven WCs (black) or fixing them to the SM value of zero (red).
  }
\label{fig:eft_results}
\end{figure}

\begin{table}
\centering
\topcaption{Observed 95\% \CL intervals for the eight WCs in the EFT model.  The intervals are determined by scanning over a single WC while either profiling the other seven WCs or fixing them to their SM value of zero.}
\renewcommand{\arraystretch}{1.4}
\begin{scotch}{ c c c }
	WC$/\Lambda^2$      & \multicolumn{2}{c}{95\% \CL interval [$\TeVns^{-2}$]} \\
                        & (Others profiled) & (Others fixed to SM)              \\
	\hline
	$\ctp  / \Lambda^2$ & [0.56, 30]        & [0.20, 30]                        \\
	$\cpQM / \Lambda^2$ & [$-8.3$, 9.9]     & [$-6.6$, 8.7]                     \\
	$\cpQa / \Lambda^2$ & [$-4.4$, 3.9]     & [$-4.1$, 3.0]                     \\
	$\cpt  / \Lambda^2$ & [$-13$, 7.9]      & [$-12$, 6.3]                      \\
	$\cptb / \Lambda^2$ & [$-10$, 12]       & [$-9.9$, 11]                      \\
	$\ctW  / \Lambda^2$ & [$-1.6$, 1.6]     & [$-1.0$, 0.96]                    \\
	$\cbW  / \Lambda^2$ & [$-4.3$, 4.3]     & [$-4.2$, 4.2]                     \\
	$\ctZ  / \Lambda^2$ & [$-1.7$, 1.7]     & [$-1.0$, 1.1]                     \\
  \end{scotch}
\label{tab:EFTresults}
\end{table}

\begin{figure}
  \centering
    \includegraphics[width=0.48\textwidth]{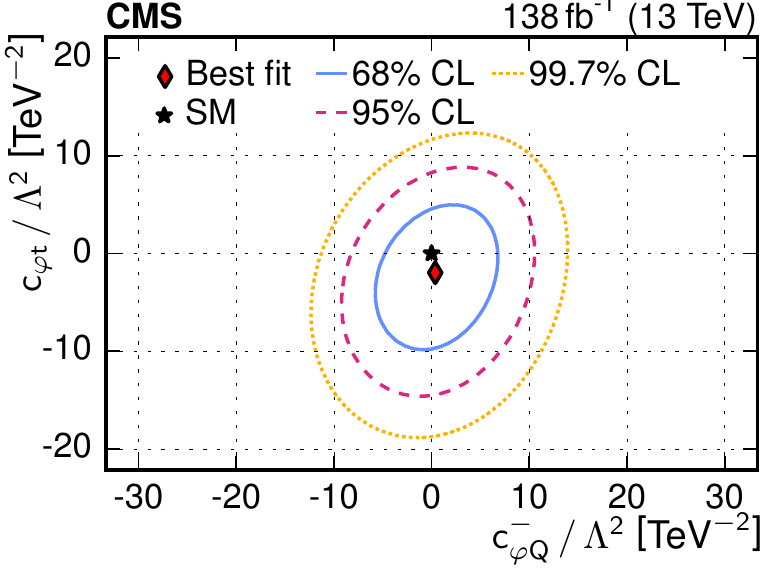}
    \includegraphics[width=0.48\textwidth]{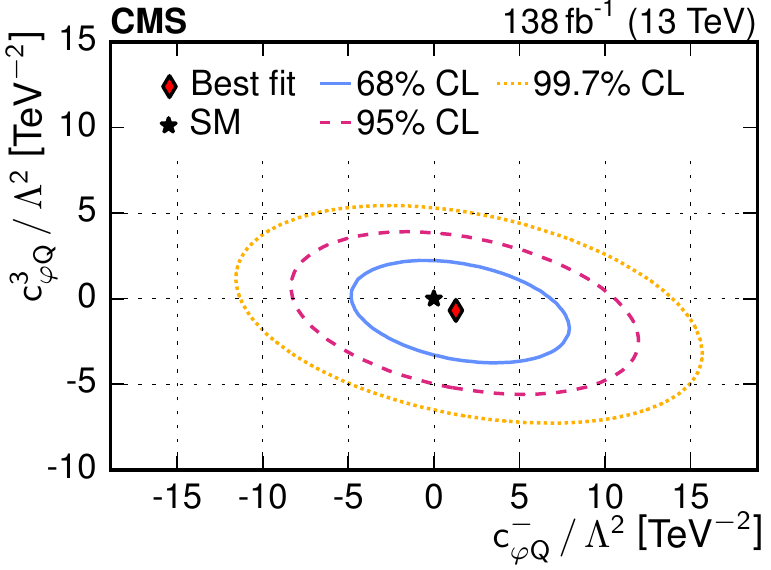}
    \includegraphics[width=0.48\textwidth]{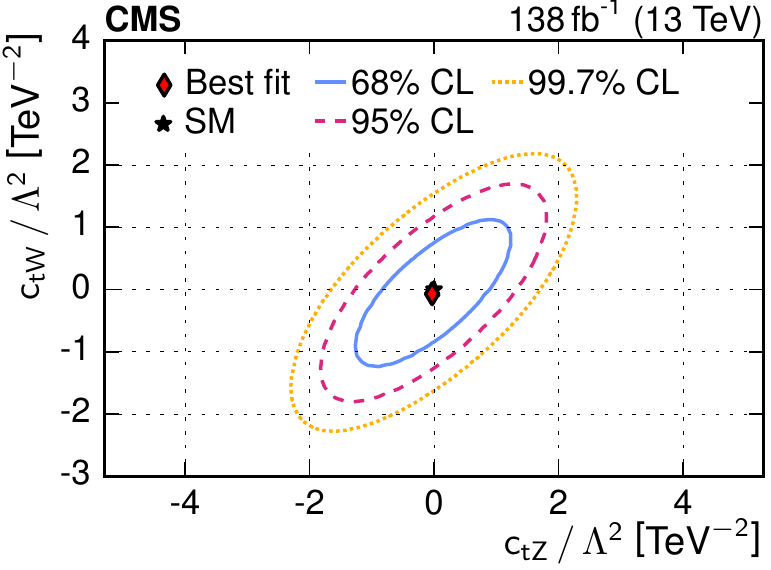}
    \caption{
    Observed two-dimensional scans of the negative log-likelihood as a function of two of the eight WCs when all other WCs are fixed to their SM values.
	The three pairs of WCs scanned have the three largest observed correlation coefficients among all pairs.
	They are \cpt versus \cpQM (\cmsUpperLeft), \cpQa versus \cpQM (\cmsUpperRight), and \ctW versus \ctZ (lower).
    The 68, 95, and 99.7\% \CL intervals are indicated by the solid blue, dashed red, and dotted orange lines, respectively.
    }
   \label{fig:EFT2dscans}
\end{figure}

The likelihood scans for the WCs are sometimes bimodal, as can be seen in Fig.~\ref{fig:EFTresults2} for \ctp and \cbW.
This arises because the predicted yields in each analysis bin are quadratic as functions of the WCs, so the observed or expected yields may be most consistent with the predictions at two distinct WC values.
The observed WC constraints are tighter than expected because the observed yields of \ttbb and \ttH are higher and lower, respectively, than the SM expectations.
All of the WC constraints are in close agreement with the SM with the exception of \ctp, which shows a weak tension with the SM.
This tension is dominated by the yield of \ttH, which is smaller than expected, as seen in Figs.~\ref{fig:signalStrengths} and \ref{fig:diffXS}.

Figure~\ref{fig:eft_summary} compares the 95\% \CL intervals from this analysis with previous CMS measurements, which used events with multiple leptons or photons~\cite{CMS-TOP-19-001,CMS-TOP-18-009,CMS-TOP-21-001,CMS-TOP-21-004}.
There is good agreement among the various results, and in some cases the limits from this analysis are comparable to the best previously published limits.

\begin{figure}
  \centering
  \includegraphics[width=\onecolwidth]{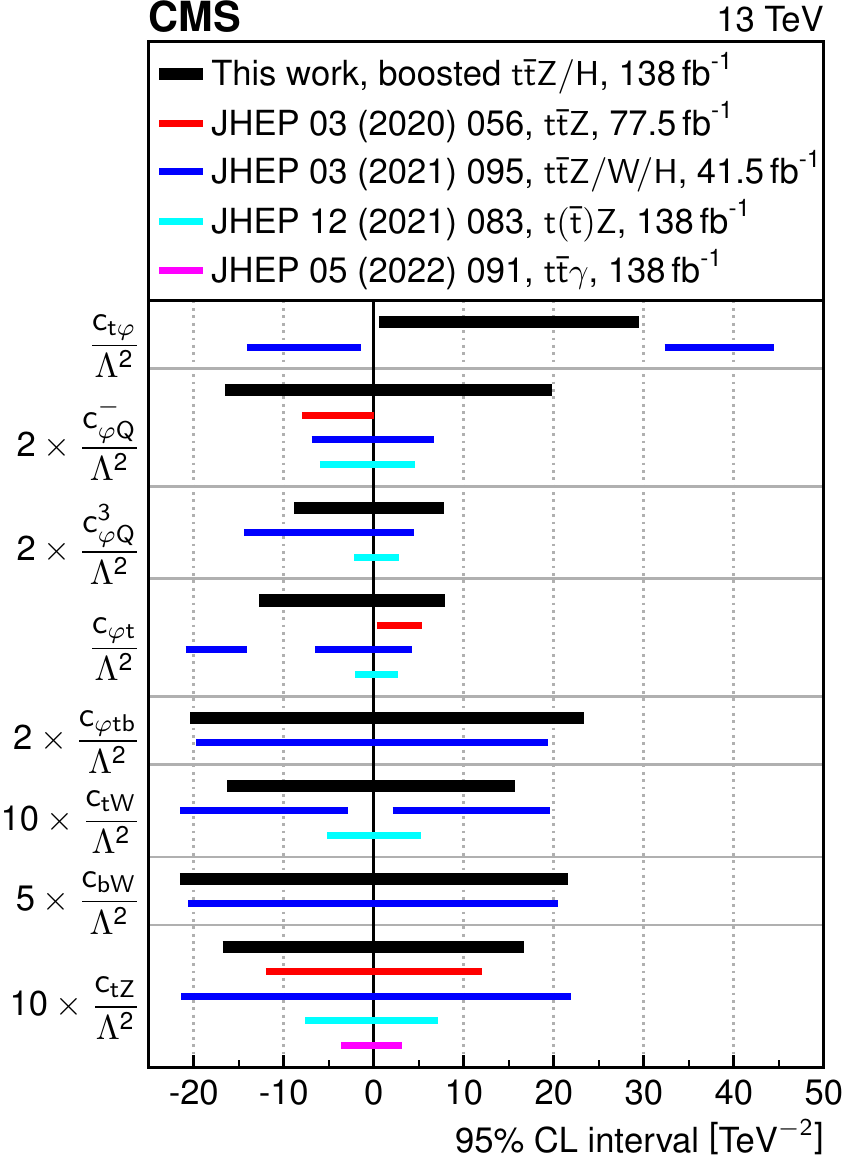}
  \caption{
	Observed 95\% \CL intervals for the WCs. The intervals are found by scanning over a single WC while fixing the other seven to zero.
	For comparison, we also show the corresponding 95\% \CL intervals from Refs.~\cite{CMS-TOP-18-009,CMS-TOP-19-001,CMS-TOP-21-001,CMS-TOP-21-004}, which used events with multiple leptons or photons.
  The intervals for several of the WCs would be too small to see clearly, and so we have increased their size by the factor given in the label on the left edge of the figure.
  }
\label{fig:eft_summary}
\end{figure}

\section{Summary}
\label{sec:summary}
Measurements of the signal strengths and 95\% confidence level upper limits on the differential cross sections for production of \ttZ and \ttH events, where \PH refers to the Higgs boson, are presented along with constraints on the Wilson coefficients of a leading-order effective field theory.
The analysis is performed using the $\bbbar$ decay mode of the \PZ or Higgs boson and the lepton plus jets channel of the associated \ttbar pair.
The \PZ or Higgs boson is required to be Lorentz boosted, with transverse momentum $\pt > 200\GeV$.
A deep neural network is employed to discriminate between the \ttZ and \ttH signal events and the background, which is dominated by \ttjets production.
The data correspond to an integrated luminosity of 138\fbinv collected with the CMS detector at the CERN LHC from 2016 through 2018.
The data are binned as a function of the deep neural network score and the reconstructed \pt and mass of the \PZ or Higgs boson.
Binned maximum likelihood fits are employed to extract the observables from the data.

The data are found to be consistent with the expectations from the standard model.
The signal strength modifiers for boosted \ttZ and \ttH production are measured to be $\mu_{\ttZ} = {0.65\,}^{+1.04}_{-0.98}$ and $\mu_{\ttH} = {-0.27\,}^{+0.86}_{-0.83}$, which are both consistent with the expected value, 1.
The 95\% confidence level upper limits on the differential \ttZ and \ttH cross sections range from 2 to 5 times the standard model predicted cross sections for \PZ or Higgs boson $\pt>300\GeV$.
Results are also presented on eight parameters of a leading-order effective field theory that have a large impact on boosted \ttZ and \ttH production.
These results represent the most restrictive limits to date on the cross sections for the production of \ttZ and \ttH with \PZ or Higgs boson $\pt > 450\GeV$.
The limits on the Wilson coefficients in the effective field theory are consistent and, in some cases, competitive with the best previous limits.

\begin{acknowledgments}
We congratulate our colleagues in the CERN accelerator departments for the excellent performance of the LHC and thank the technical and administrative staffs at CERN and at other CMS institutes for their contributions to the success of the CMS effort. In addition, we gratefully acknowledge the computing centers and personnel of the Worldwide LHC Computing Grid and other centers for delivering so effectively the computing infrastructure essential to our analyses. Finally, we acknowledge the enduring support for the construction and operation of the LHC, the CMS detector, and the supporting computing infrastructure provided by the following funding agencies: BMBWF and FWF (Austria); FNRS and FWO (Belgium); CNPq, CAPES, FAPERJ, FAPERGS, and FAPESP (Brazil); MES and BNSF (Bulgaria); CERN; CAS, MoST, and NSFC (China); MINCIENCIAS (Colombia); MSES and CSF (Croatia); RIF (Cyprus); SENESCYT (Ecuador); MoER, ERC PUT and ERDF (Estonia); Academy of Finland, MEC, and HIP (Finland); CEA and CNRS/IN2P3 (France); BMBF, DFG, and HGF (Germany); GSRI (Greece); NKFIH (Hungary); DAE and DST (India); IPM (Iran); SFI (Ireland); INFN (Italy); MSIP and NRF (Republic of Korea); MES (Latvia); LAS (Lithuania); MOE and UM (Malaysia); BUAP, CINVESTAV, CONACYT, LNS, SEP, and UASLP-FAI (Mexico); MOS (Montenegro); MBIE (New Zealand); PAEC (Pakistan); MES and NSC (Poland); FCT (Portugal); MESTD (Serbia); MCIN/AEI and PCTI (Spain); MOSTR (Sri Lanka); Swiss Funding Agencies (Switzerland); MST (Taipei); MHESI and NSTDA (Thailand); TUBITAK and TENMAK (Turkey); NASU (Ukraine); STFC (United Kingdom); DOE and NSF (USA).
	
\hyphenation{Rachada-pisek} Individuals have received support from the Marie-Curie program and the European Research Council and Horizon 2020 Grant, contract Nos.\ 675440, 724704, 752730, 758316, 765710, 824093, 884104, and COST Action CA16108 (European Union); the Leventis Foundation; the Alfred P.\ Sloan Foundation; the Alexander von Humboldt Foundation; the Belgian Federal Science Policy Office; the Fonds pour la Formation \`a la Recherche dans l'Industrie et dans l'Agriculture (FRIA-Belgium); the Agentschap voor Innovatie door Wetenschap en Technologie (IWT-Belgium); the F.R.S.-FNRS and FWO (Belgium) under the ``Excellence of Science -- EOS" -- be.h project n.\ 30820817; the Beijing Municipal Science \& Technology Commission, No. Z191100007219010; the Ministry of Education, Youth and Sports (MEYS) of the Czech Republic; the Hellenic Foundation for Research and Innovation (HFRI), Project Number 2288 (Greece); the Deutsche Forschungsgemeinschaft (DFG), under Germany's Excellence Strategy -- EXC 2121 ``Quantum Universe" -- 390833306, and under project number 400140256 - GRK2497; the Hungarian Academy of Sciences, the New National Excellence Program - \'UNKP, the NKFIH research grants K 124845, K 124850, K 128713, K 128786, K 129058, K 131991, K 133046, K 138136, K 143460, K 143477, 2020-2.2.1-ED-2021-00181, and TKP2021-NKTA-64 (Hungary); the Council of Science and Industrial Research, India; the Latvian Council of Science; the Ministry of Education and Science, project no. 2022/WK/14, and the National Science Center, contracts Opus 2021/41/B/ST2/01369 and 2021/43/B/ST2/01552 (Poland); the Funda\c{c}\~ao para a Ci\^encia e a Tecnologia, grant CEECIND/01334/2018 (Portugal); the National Priorities Research Program by Qatar National Research Fund; MCIN/AEI/10.13039/501100011033, ERDF ``a way of making Europe", and the Programa Estatal de Fomento de la Investigaci{\'o}n Cient{\'i}fica y T{\'e}cnica de Excelencia Mar\'{\i}a de Maeztu, grant MDM-2017-0765 and Programa Severo Ochoa del Principado de Asturias (Spain); the Chulalongkorn Academic into Its 2nd Century Project Advancement Project, and the National Science, Research and Innovation Fund via the Program Management Unit for Human Resources \& Institutional Development, Research and Innovation, grant B05F650021 (Thailand); the Kavli Foundation; the Nvidia Corporation; the SuperMicro Corporation; the Welch Foundation, contract C-1845; and the Weston Havens Foundation (USA).
\end{acknowledgments}

\bibliography{auto_generated}
\clearpage
\appendix
\numberwithin{table}{section}
\numberwithin{figure}{section}
\ifthenelse{\boolean{cms@external}}{\onecolumngrid}{}
\section{Neural network input variables}
\label{app:mva_vars}
Table~\ref{tab:inputvariables} lists the input variables of the DNN.
The distributions of six of these input variables are shown in Fig.~\ref{fig:nn_input_validation}.

\begin{table*}[hbp]
	\centering
	\topcaption{
		Comprehensive list of the input variables of the DNN, which is described in Section~\ref{sec:discrimination}. The ``+'' represents the relativistic four-momentum sum. Some variables are calculated for both the highest \pt (leading) and second-highest \pt (subleading) jet as indicated.
	}
	\label{tab:inputvariables}
	\cmsTable{
	\begin{scotch}{lp{\apptabwidth}}

		Name & Description \\
		\hline \\[-2ex]
		\multicolumn{2}{l}{\textbf{\ttbar system}}\\

		\PQb \pt   & \pt of the leading (subleading) \PQb jet\\
		\PQb score & \textsc{DeepCSV} score of the leading (subleading) \PQb jet\\
		\PQq \pt     & \pt of the leading (subleading) non-\PQb jet\\
		\PQq score   & \textsc{DeepCSV} score of the leading (subleading) non-\PQb jet\\
		$\DR(\PQb, \PQq)$  & minimum \DR{} between the leading (subleading) \PQb jet and any non-\PQb jet\\
		$\DR(\PQq, \PQq)$ & \DR{} between the two non-\PQb jets closest to the leading (subleading) \PQb jet\\
		$m(\PQq+\PQq)$ & invariant mass of the two non-\PQb jets closest to the leading (subleading) \PQb jet\\
		$\DR(\PQb, \PQq+\PQq)$ & \DR{} between the leading (subleading) \PQb jet and the sum of the nearest two non-\PQb jets \\
		$m(\PQb+\PQq+\PQq)$   & invariant mass of the leading (subleading) \PQb jet and the nearest two non-\PQb jets\\
		$\DR(\PZ/\PH, \PQb+\PQq+\PQq)$ & \DR{} between the \PZ or Higgs boson candidate and the sum of the leading (subleading) \PQb jet and the two non-\PQb jets nearest to the leading (subleading) \PQb jet \\
		$\DR(\PZ/\PH, \PQb+\PQb+\PQq+\PQq+\Pell)$ & \DR{} between the \PZ or Higgs boson candidate and the sum of the leading and subleading \PQb jets, the two non-\PQb jets nearest to the leading (subleading) \PQb jet, and the lepton\\
		$\mT(\PQb+\Pell, \ptvecmiss)$ & transverse mass of the subleading \PQb jet and the lepton \\
		$m(\PZ/\PH+\PQb)$   & invariant mass of the \PZ or Higgs boson candidate and the nearest \PQb jet \\
		$m(\PQb+\PQb)$        & invariant mass of the leading and subleading \PQb jets \\
		$\DR(\PQb, \PQb)$      & \DR{} between the leading and subleading \PQb jets \\
		$\DR(\PZ/\PH, \PQq)$      & \DR{} between the \PZ or Higgs boson candidate and the leading non-\PQb jet \\
		$\DR(\PZ/\PH, \PQb)$      & \DR{} between the \PZ or Higgs boson candidate and the leading \PQb jet \\
		$\DR(\PZ/\PH, \Pell)$  & \DR{} between the \PZ or Higgs boson candidate and the lepton\\
		$m(\PZ/\PH+\Pell)$ & invariant mass of the \PZ or Higgs boson candidate and the lepton\\
		$\DR(\PQb, \Pell)$       & \DR{} between the leading (subleading) \PQb jet and the lepton\\
		$m(\PQb+\Pell)$ & invariant mass of the leading (subleading) \PQb jet and the lepton\\
		$N(\PQb_\text{out})$       & number of \PQb jets outside the \PZ or Higgs boson candidate cone (${\DR > 0.8}$) \\
		$N(\PQq_\text{out})$       & number of non-\PQb jets outside the \PZ or Higgs boson candidate cone (${\DR > 0.8}$) \\ [\cmsTabSkip]

		\multicolumn{2}{l}{\textbf{Event topology}}\\

		$N(\text{AK8 jets})$         & number of AK8 jets, including the \PZ or Higgs boson candidate\\
		$N(\text{AK4 jets})$         & number of AK4 jets\\
		$N(\PZ/\PH)$       & number of AK8 jets with a minimum AK8 $\bbbar$ tagger score of 0.8\\
		AK8 \msd & maximum \msd of AK8 jets, excluding the \PZ or Higgs boson candidate \\
		$\HT(\PQb_\text{out})$ & \HT of the \PQb jets outside the \PZ or Higgs boson candidate cone (${\DR > 0.8}$)\\
		$\HT(\PQb_\text{out},\PQq_\text{out},\Pell)$  & \HT of all AK4 jets outside the \PZ or Higgs boson candidate cone (${\DR > 0.8}$) and the lepton\\
		sphericity  & sphericity calculated from the AK4 jets and the lepton~\cite{spher_aplan} \\
		aplanarity & aplanarity calculated from the AK4 jets and the lepton~\cite{spher_aplan}\\ [\cmsTabSkip]

		\multicolumn{2}{l}{\textbf{\PZ or Higgs boson candidate substructure}}\\

		$\PQb_\text{in}$ score     & maximum (minimum) \textsc{DeepCSV} score of AK4 jets within the \PZ or Higgs boson candidate cone (${\DR < 0.8}$)\\
		$\DR(\PQb_\text{in}, \PQb_\text{out})$  & \DR{} between one \PQb jet within the \PZ or Higgs boson candidate cone (${\DR < 0.8}$) and the leading \PQb jet outside of the \PZ or Higgs boson candidate cone\\
		$N(\PQb_\text{in})$         & number of \PQb jets within the \PZ or Higgs boson candidate cone (${\DR < 0.8}$) \\
		$N(\PQq_\text{in})$         & number of non-\PQb jets within the \PZ or Higgs boson candidate cone (${\DR < 0.8}$) \\
		\PZ/\PH $\bbbar$ score & AK8 $\bbbar$ tagger score of the \PZ or Higgs boson candidate \\

	\end{scotch}
	}
\end{table*}

\begin{figure*}
  \centering
  \begin{tabular}{cc}
    \includegraphics[width=0.415\textwidth]{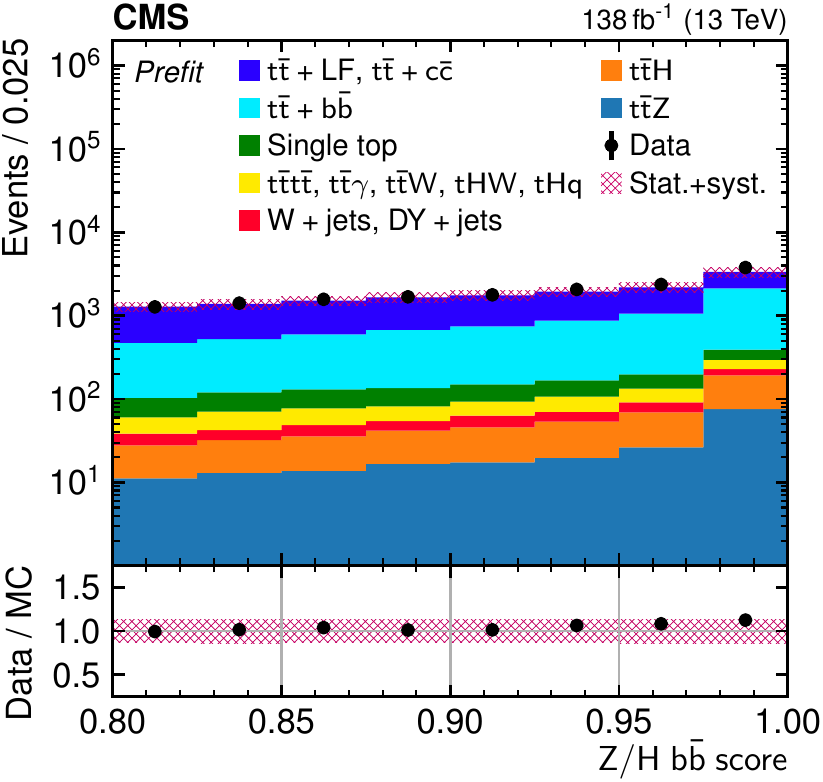} &
    \includegraphics[width=0.415\textwidth]{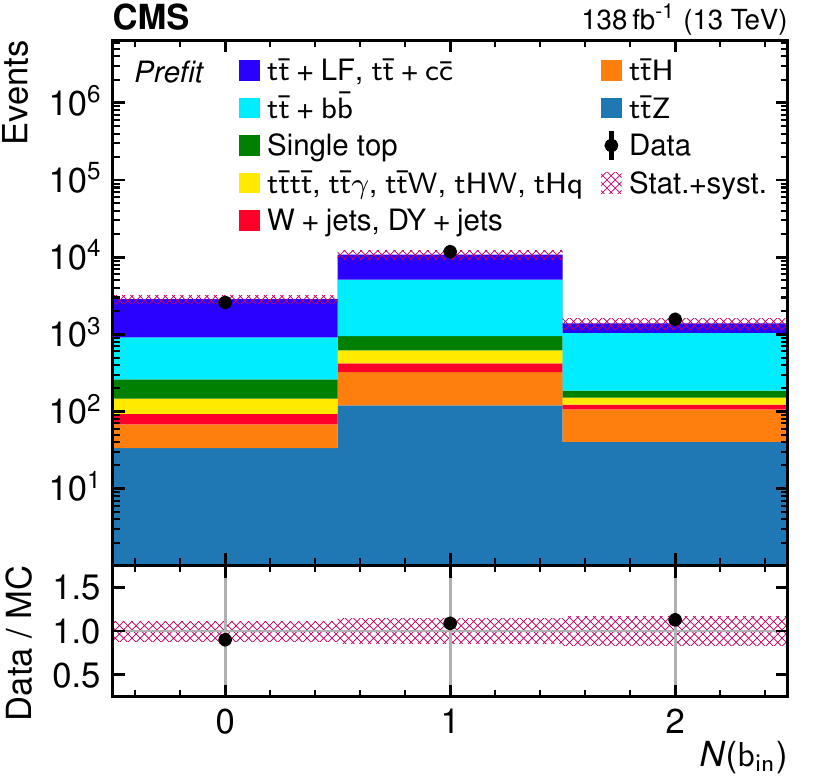} \\
    \includegraphics[width=0.415\textwidth]{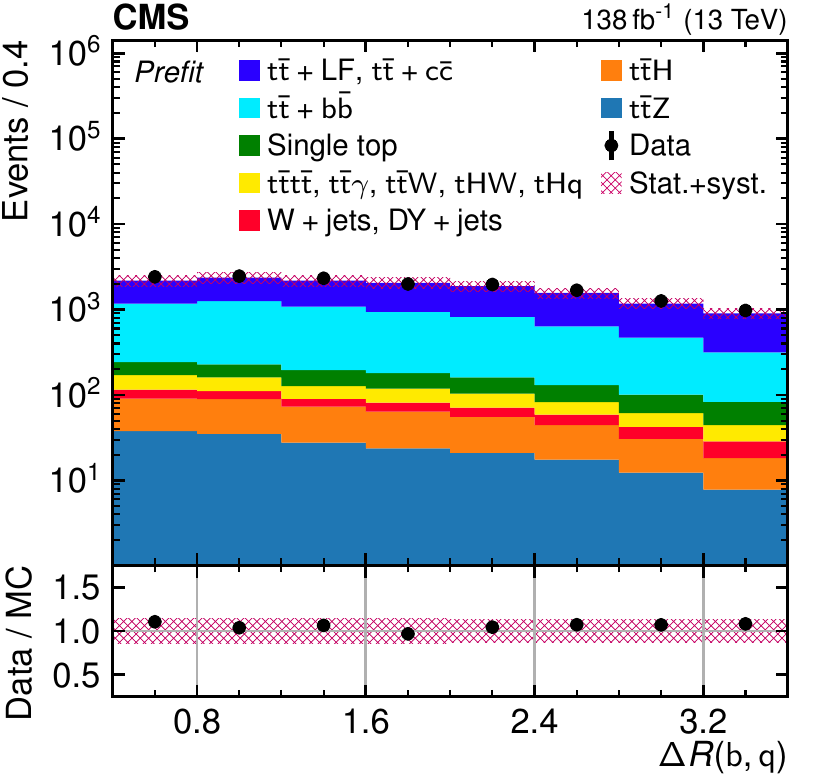} &
    \includegraphics[width=0.415\textwidth]{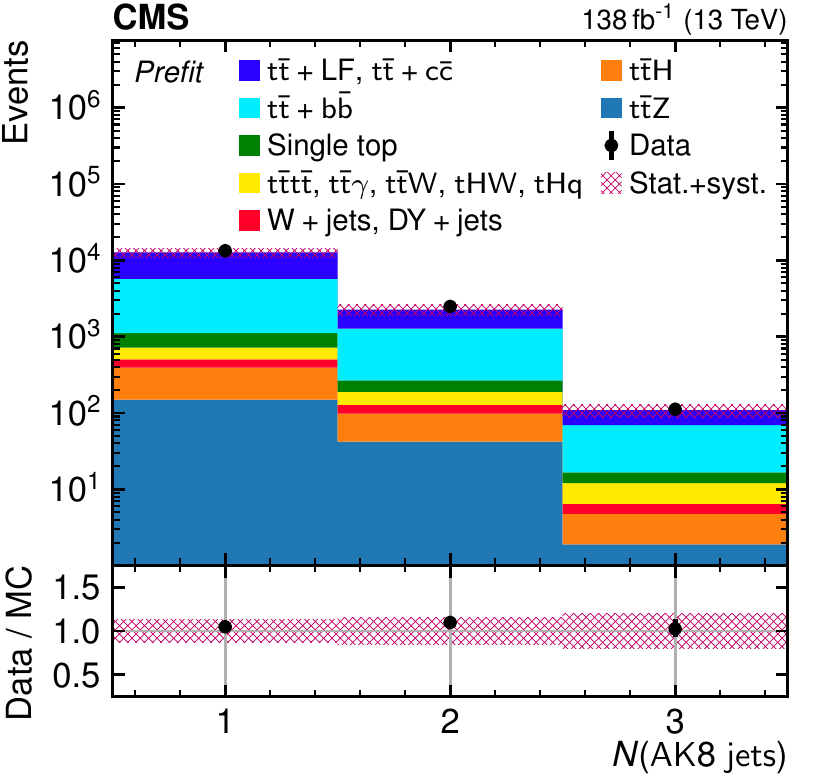} \\
    \includegraphics[width=0.415\textwidth]{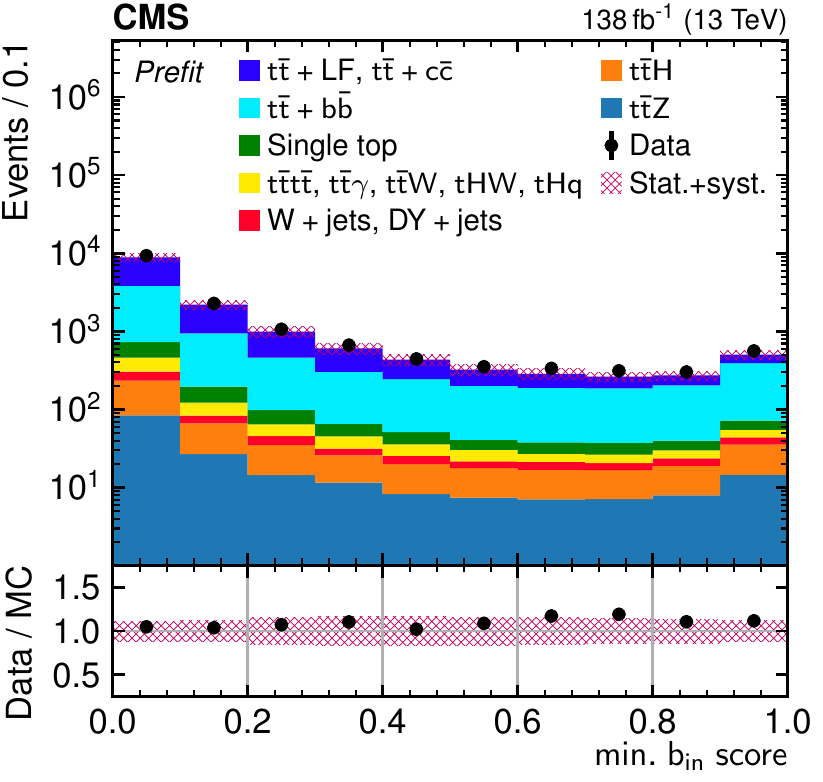} &
    \includegraphics[width=0.415\textwidth]{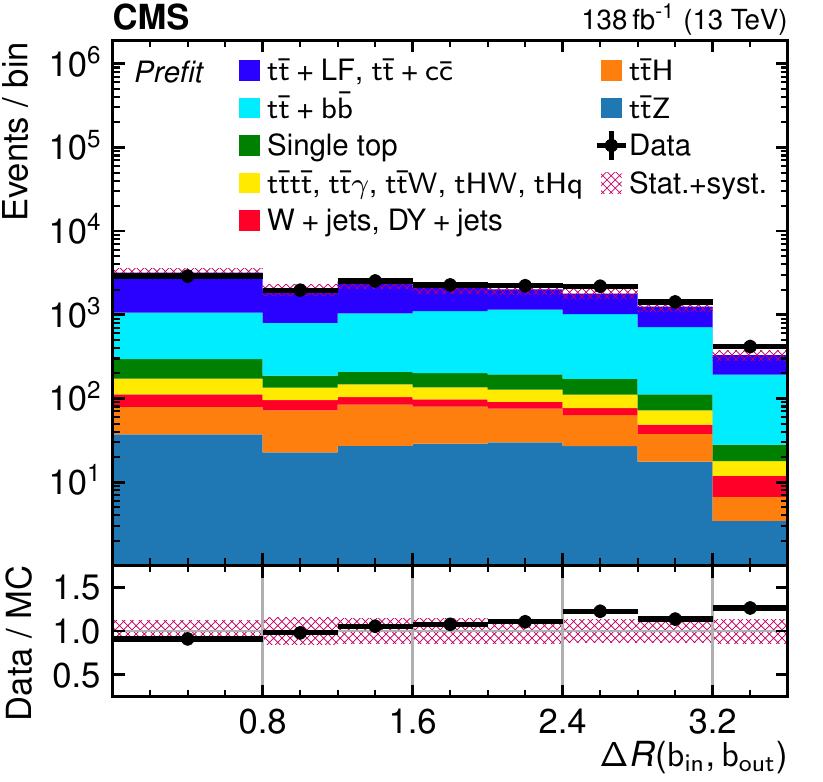} \\
  \end{tabular}
    \caption{
    	Prefit expected (colored histograms) and observed (point) distributions of selected DNN input variables, each of which is described in detail in Table~\ref{tab:inputvariables}.
		These distributions represent the combined 2016--2018 data-taking periods.
		The hatched bands represent the total statistical and systematic uncertainty in the expected distributions, while the vertical bars on the black points indicate the statistical uncertainty in the observed distributions and the horizontal bars indicate the bin widths.
		The lower panels show the ratio of the observed yields to the sum of the MC predictions.
    }
   \label{fig:nn_input_validation}
\end{figure*}

\cleardoublepage \section{The CMS Collaboration \label{app:collab}}\begin{sloppypar}\hyphenpenalty=5000\widowpenalty=500\clubpenalty=5000
\cmsinstitute{Yerevan Physics Institute, Yerevan, Armenia}
{\tolerance=6000
A.~Tumasyan\cmsAuthorMark{1}\cmsorcid{0009-0000-0684-6742}
\par}
\cmsinstitute{Institut f\"{u}r Hochenergiephysik, Vienna, Austria}
{\tolerance=6000
W.~Adam\cmsorcid{0000-0001-9099-4341}, J.W.~Andrejkovic, T.~Bergauer\cmsorcid{0000-0002-5786-0293}, S.~Chatterjee\cmsorcid{0000-0003-2660-0349}, K.~Damanakis\cmsorcid{0000-0001-5389-2872}, M.~Dragicevic\cmsorcid{0000-0003-1967-6783}, A.~Escalante~Del~Valle\cmsorcid{0000-0002-9702-6359}, P.S.~Hussain\cmsorcid{0000-0002-4825-5278}, M.~Jeitler\cmsAuthorMark{2}\cmsorcid{0000-0002-5141-9560}, N.~Krammer\cmsorcid{0000-0002-0548-0985}, L.~Lechner\cmsorcid{0000-0002-3065-1141}, D.~Liko\cmsorcid{0000-0002-3380-473X}, I.~Mikulec\cmsorcid{0000-0003-0385-2746}, P.~Paulitsch, F.M.~Pitters, J.~Schieck\cmsAuthorMark{2}\cmsorcid{0000-0002-1058-8093}, R.~Sch\"{o}fbeck\cmsorcid{0000-0002-2332-8784}, D.~Schwarz\cmsorcid{0000-0002-3821-7331}, S.~Templ\cmsorcid{0000-0003-3137-5692}, W.~Waltenberger\cmsorcid{0000-0002-6215-7228}, C.-E.~Wulz\cmsAuthorMark{2}\cmsorcid{0000-0001-9226-5812}
% ...
\par}
\vskip\cmsinstskip
\dag:~Deceased\\
$^{1}$Also at Yerevan State University, Yerevan, Armenia\\
$^{2}$Also at TU Wien, Vienna, Austria\\
% ...
\end{sloppypar}
\end{document}